\documentclass[acmsmall,screen,nonacm]{acmart}

\usepackage{booktabs}             

\usepackage{soul}
\usepackage{listings}
\lstdefinelanguage{json}{}
\usepackage{subcaption}
\usepackage{multirow}
\usepackage{float}
\usepackage{array}
\usepackage[table]{xcolor}
\usepackage{colortbl}

\newcommand{\rev}[1]{#1}
\newcommand{\revnew}[1]{#1}

\usepackage{graphicx}
\graphicspath{{./}}

\acmJournal{PACMHCI}
\acmYear{2026}
\acmVolume{0}
\acmNumber{CSCW}
\acmArticle{0}
\acmMonth{0}
\setcopyright{none}
\begin{document}

\title{Choose Your Agent: Tradeoffs in Adopting AI Advisors, Coaches, and Delegates in Multi-Party Negotiation}

\author{Kehang Zhu}
\authornote{Work done during a student researcher internship at Google DeepMind. The authors thanks David Parkes, John Horton, Michael Terry, Hal Abelson, Benjamin Manning, Peyman Shahidi, Anand Shah, Gili Rusak and participants at the Harvard EconCS Seminar for helpful comments and discussions.}
\email{kehangzhu@gmail.com}
\affiliation{%
  \institution{Harvard University}
  \city{Cambridge}
  \country{United States}
}
\author{Nithum Thain}
\affiliation{%
  \institution{Google DeepMind}
  \country{Canada}
}
\author{Vivian Tsai}
\affiliation{%
  \institution{Google DeepMind}
  \country{United States}
}
\author{James Wexler}
\affiliation{%
  \institution{Google DeepMind}
  \country{United States}
}
\author{Crystal Qian}
\affiliation{%
  \institution{Google DeepMind}
  \city{New York}
  \country{United States}
}

\renewcommand{\shortauthors}{Zhu et al.}

\begin{abstract}
As AI usage becomes more prevalent in social contexts, understanding agent-user interaction is critical to designing systems that improve both individual and group outcomes. We present an online behavioral experiment ($N=243$) in which participants play three multi-turn bargaining games in groups of three.
Each game, presented in randomized order, grants \textit{access to} a single LLM assistance modality: proactive recommendations from an \textit{Advisor}, reactive feedback from a \textit{Coach}, or autonomous execution by a \textit{Delegate}. All three modalities are powered by an LLM with \rev{super-human performance within this negotiation setting}.
On each turn, participants privately decide whether to act manually or use the AI modality available in that game.
We document a \emph{preference--performance misalignment}: participants strongly prefer the higher-control \textit{Advisor} (44\%) over the \textit{Delegate} (19\%), yet groups \rev{only significantly increase collective surplus under Delegate access}. Adjusting for voluntary non-compliance, delegating to the AI yields suggestive individual welfare gains, roughly 1.5$\times$ the intent-to-treat estimate.
A mechanism analysis traces this gap to a \emph{human filter}: AI-generated proposals create more joint surplus than manual proposals across all conditions, but in the Advisor and Coach modes users modify, override, or ignore the AI's suggestions, reverting toward human-baseline trade patterns. The Delegate advantage arises not from a different AI capability but from \rev{bypassing this filtering step altogether}. \rev{Realizing these welfare gains depends not only on model capability, but on the interaction structure through which that capability is delivered.} We argue that assistance modalities should be designed as mechanisms with endogenous participation; adoption-compatible interaction rules are a prerequisite to improving welfare with automated assistance.
\end{abstract}

\ccsdesc[500]{Human-centered computing~Empirical studies in collaborative and social computing}
\ccsdesc[300]{Human-centered computing~Empirical studies in HCI}
\ccsdesc[300]{Computing methodologies~Multi-agent systems}

\keywords{human-AI interaction; large language models; AI delegation;
  negotiation; behavioral experiment; collaborative decision making}

\maketitle

\section{Introduction}

Large language models (LLMs) are shifting from passive tools to autonomous \emph{agents}, capable of navigating complex social tasks and reshaping the incentives of multi-party environments. Understanding the design and effects of such systems is \rev{key to} the emerging ``agentic economy''~\cite{tomasev2025virtual, rothschild2025agentic, shahidi2025coasean}; \rev{furthermore, in shared environments, how much agency each stakeholder cedes to an AI directly shapes collective welfare alongside individual outcomes.}

The tension \rev{in delegating actions to autonomous agents} is increasingly operationalized through a diverse spectrum of assistance modalities. Beyond traditional chat interfaces that passively respond to user input, emerging systems range from proactive assistants that initiate guidance~\cite{openai_study_mode_2025, google_guided_learning_2025} to autonomous agents empowered with independent execution~\cite{openai_chatgpt_agent_2025}. However, the impact of these design choices on strategic outcomes—and their scalability in multi-party environments—remains \rev{underexplored}.

Prior research on human-AI collaboration has primarily evaluated single-user, single-AI interactions in domains such as medical diagnosis, credit assessment, and risk prediction~\cite{agarwal2023combining, green2019principles, bansal2019beyond}, often targeting objectives with verifiable ground truths~\cite{vaccaro2024combinations}. These studies have established foundational insights, such as a preference-performance misalignment, where users may prefer sub-optimal agents that require less cognitive load~\cite{bansal2019beyond, bucinca2021trust, bucinca2020proxy}.
However, virtually all empirical evidence on human--AI interaction comes from \emph{individual} decision-making tasks---settings where one user interacts with one AI system and the outcome depends only on that dyad. Experimental data on AI assistance in \emph{multi-party, strategically interdependent} environments---where one participant's AI-assisted action changes the payoff landscape for others---remains scarce.
This gap matters because multi-party settings introduce externalities, equilibrium effects, and adoption spillovers that cannot be studied in isolated dyads. \rev{Understanding these dynamics requires moving beyond individual users to examine how AI-mediated technology reshapes collaborative group outcomes~\cite{grudin1988cscw_fail}.}

Such human-AI interaction patterns become increasingly complex in social, real-world applications, which are defined by strategic interdependence, dynamic equilibria, and collective externalities. Recent work on autonomous LLM negotiations demonstrate that agent-to-agent bargaining can exhibit distinct risks and behaviors relative to human negotiation~\cite{zhu2025automated, qian2025strategic}. A complementary line of research studies \textit{principal-agent} interaction patterns in social contexts, where humans customize an agent (often via prompt writing) to negotiate autonomously on their behalf~\cite{imas2025agentic, vaccaro2025advancing}. While these studies establish the feasibility of autonomous negotiation, they abstract away from the practical design choice of user controllability---specifically, how different modalities of interaction influence adoption and equilibrium outcomes.\footnote{For example, coding agents like Claude-Code offer three distinct modes of assistance: \textit{planning mode}, \textit{ask before edits}, and \textit{edit automatically}.}

We study how different allocations of agency affect behavior and welfare in a strategically interdependent setting, operationalizing the spectrum of AI agency into three distinct interaction modalities while holding the underlying model capability constant: an \textit{Advisor} (proactive recommendations), a \textit{Coach} (reactive feedback), and a \textit{Delegate} (autonomous action).

These three modalities are grounded in the levels-of-automation framework proposed by \citet{sheridan1978supervisory} and formalized by \citet{parasuraman2000model}, who characterize a continuum from full human control to full machine autonomy. Rather than sampling arbitrarily along this continuum, our design selects three qualitatively distinct allocations of decision authority: the Advisor initiates action and the human retains veto power; the Coach inverts this flow, with the human initiating and the system critiquing before execution; and the Delegate acts autonomously on the human's behalf.
These three points represent the principal ways in which initiative and oversight can be partitioned between a human and an AI assistant: AI-proposes/human-decides, human-proposes/AI-critiques, and AI-decides/human-observes.\footnote{These modalities have close analogues in deployed AI products. The Advisor mode corresponds to the default interaction pattern of most commercial LLMs, in which the system proactively generates a response that the user can accept, edit, or discard. The Coach mode mirrors guided-learning features such as Google's Guided Learning in Gemini~\cite{google_guided_learning_2025} and OpenAI's Study Mode~\cite{openai_study_mode_2025}, which prompt the user to reason first and then provide targeted feedback. The Delegate mode corresponds to the autonomous agent products now offered by all major AI providers---for example, OpenAI's ChatGPT Agent~\cite{openai_chatgpt_agent_2025} and Google's Gemini Agent Mode~\cite{google_gemini_agent_mode_2025}---in which the system takes actions on the user's behalf without requiring approval at each step.}

We evaluate these modalities through a bargaining game~\cite{qian2025strategic} designed to evaluate AI capabilities in group negotiations, a setting with empirically exhibited human inefficiencies in trading behaviors~\cite{rubinstein1982perfect, binmore1986nash}. Our LLM agents utilize prompt scaffolding based on \texttt{Gemini-2.5 Flash}~\cite{comanici2025gemini} that \rev{empirically outperforms human traders in the All-Agent Baseline}, allowing us to isolate the effects of the interaction structure. We then conduct a randomized, within-subjects experiment ($N=243$) where participants engage in three successive games, using each modality in a counterbalanced order. \rev{Because each game involves three interdependent players, one participant's AI use directly affects the proposals and welfare available to counterparts.}

\rev{Our study provides empirical evidence for the following:}
\begin{itemize}
    \item \rev{\textbf{Preference--performance misalignment and the human filter.}
    In a within-subject experiment ($N=243$), participants strongly prefer their games with access to the Advisor modality, yet groups achieve the highest collective surplus with the autonomous Delegate. A mechanism analysis reveals that human intervention in Advisor and Coach modes dilutes the quality of the offers---users override, modify, or ignore AI suggestions to align them with risk-averse social norms, reducing the welfare benefits of the AI-generated offers.}
    
    \item \rev{\textbf{Market-making spillovers in shared environments.}
    We demonstrate that Delegate adoption creates robust positive externalities. Rather than exploiting unassisted counterparts to capture surplus, autonomous agents act as cooperative market-makers. By introducing high-quality proposals, they structurally upgrade the shared offer pool and increase surplus for unassisted counterparts by $21.6\%$.}
\end{itemize}

Although the LLM agents \rev{outperform human traders in the All-Agent Baseline}, our randomized study shows that users do not reliably \rev{adopt them} due to interaction frictions. This suggests that providing \rev{strong} agentic capabilities alone \rev{is} insufficient for improving human outcomes; interface design and interaction patterns are central to realizing benefits in collective systems.

We emphasize that our findings are established within a specific, stylized bargaining game; whether the same patterns hold in other strategic domains---with different payoff structures, communication channels, or time horizons---remains an open empirical question. Nevertheless, by documenting the preference--performance misalignment and the human-filter mechanism with clean causal identification, our work provides an early baseline and a reusable experimental paradigm for studying human--agent interaction in strategic, multi-party ecosystems. \rev{We conclude with design implications for collaborative AI systems---including delegation with veto windows, progressive confidence disclosure, and group-level evaluation frameworks---grounded in our empirical findings.}

\begin{figure}
  \includegraphics[width=\textwidth]{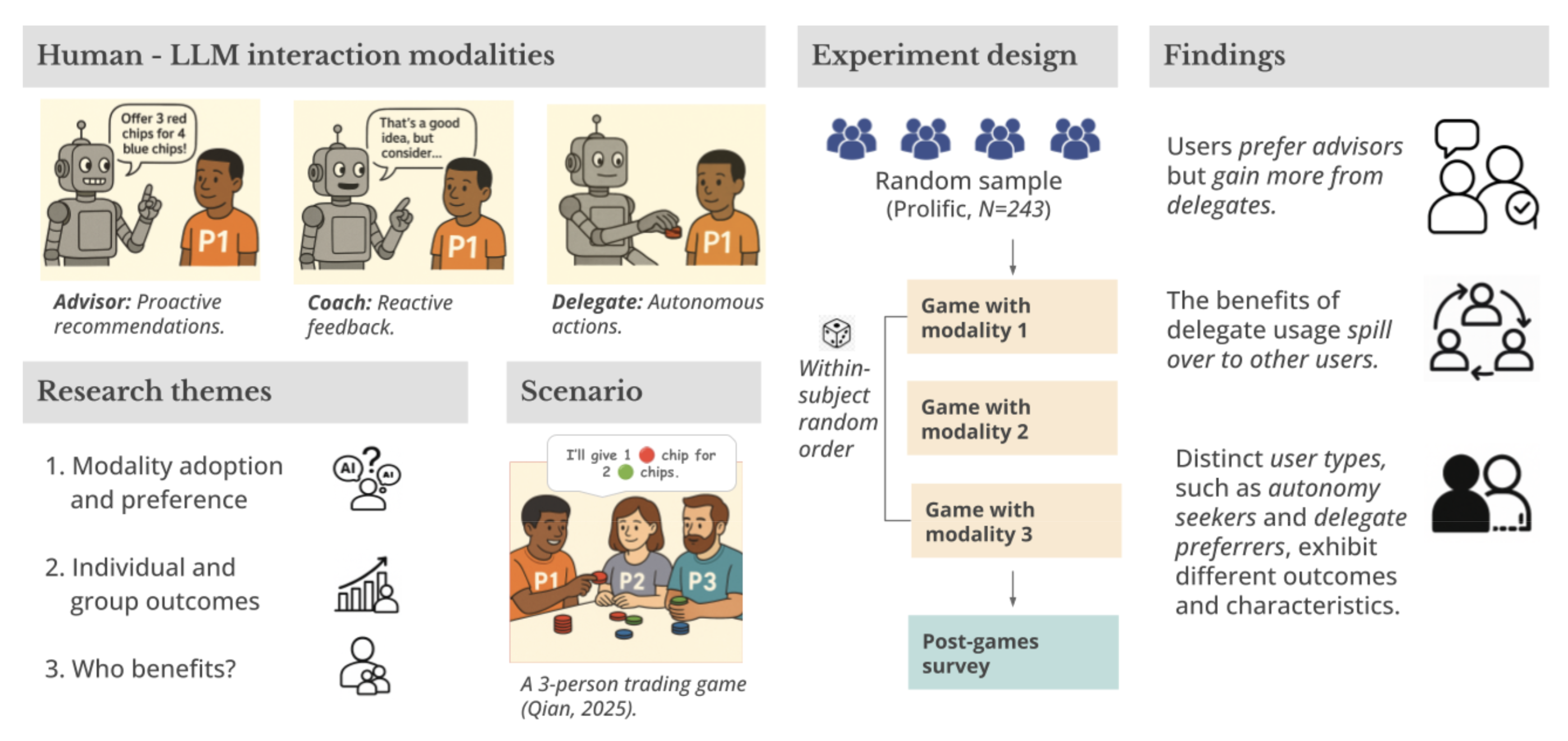}
  \caption{Overview of the experimental design and contributions. Participants ($N=243$) engaged in three-person bargaining games with access to three LLM assistance modalities: \textit{Advisor} (proactive recommendations), \textit{Coach} (reactive feedback), or \textit{Delegate} (autonomous actions)---with within-subject game randomization. We document preference--performance misalignment (users prefer Advisor but achieve highest welfare with Delegate) and identify a \emph{human filter} mechanism: in Advisor and Coach modes, users modify, override, or ignore high-quality AI proposals, diluting their welfare benefit.}
  \label{fig:teaser}
\end{figure}

\section{Related Work}

\subsection{Negotiation and Group Decision Making}

LLM-based agents demonstrate increasingly strong social capabilities across negotiation, persuasion, mediation, and multi-agent coordination tasks~\cite{horton2023large,manning2024automated,shah2025learning,abdelnabi2023llm,simmsai,tessler2024ai,agashe2023llm,li2023metaagents,soumalias2025llm}. Modern LLMs \rev{achieve} human-level strategic performance in complex multi-party settings such as \emph{Diplomacy} \cite{meta2022human}, outperform human debaters in randomized tournaments \cite{simmsai}, and even produce moral judgments preferred by human evaluators \cite{palminteri_garcia_qian_2025}.


Beyond individual capabilities, LLMs have been deployed to support collective processes, such as facilitation, moderation, and mediation, in committees, communities, and group deliberation \cite{de2025amplifying,alsobay2025bringing,domingos2021delegation,tessler2024ai,kobiella2025ai}. \rev{These modern applications build on early systems for Group Decision Support (GDSS) and Negotiation Support (NSS), which demonstrated that structured communication and decision-analytic scaffolding can improve group outcomes and consensus~\cite{desanctis1987gdss, lim1992nss, kraemer1990gdss}. While early systems relied on passive optimization or rigid rules, modern generative agents act as active, conversational participants, which enable agents to provide dynamic, text-based coaching or advising. We extend this research by studying how varying the allocation of decision initiative between humans and agents shapes both adoption behavior and group welfare in resource bargaining.}

Most human-AI studies evaluate individual users (i.e. 1:1 human-AI dyads) in tasks with ground-truth data, such as diagnosis, lending, or deception detection \cite{agarwal2023combining,agarwal2025designing,green2019principles,hoong2025improving,lai2020chicago,lai2023towards}. 
Moving from dyads to multi-party bargaining introduces additional complexities as one participant’s actions change the opportunity set faced by others, creating structural externalities.

To enable controlled comparison in such environments, \citet{qian2025strategic} propose a stylized multi-party bargaining game with induced values, abstracting away subjective goals \cite{curhan2006people} while preserving strategic interdependence.  Related work also investigates fully autonomous LLM negotiators through prompt-designed competitions \cite{vaccaro2025advancing,imas2025agentic} and agent–agent market simulations \cite{zhu2025automated}. While these studies primarily evaluate agent capabilities in bargaining environments, our study evaluates the human adoption of such capabilities through varying interaction modalities.



\subsection{Delegation in Human--AI Interaction}

Full delegation to agentic systems remains relatively novel in empirical studies of human--AI teaming; a recent meta-analysis \cite{vaccaro2024combinations} reviewed over 100 experiments and found that only a small subset involve structured delegation of decision authority. When delegation is studied, it typically appears in isolated tasks where complementarity can be engineered; for example, hybrid human–AI assessments in physical therapy \cite{lee2021human}, post-editing pipelines in summarization \cite{lai2022exploration}, or selective triage allocations based on uncertainty or rule-based heuristics \cite{raghu2019automation,agarwal2025sufficient}.

In group settings, agent delegation can increase cooperation in collective-risk dilemmas \cite{domingos2021delegation}. AI-mediated group consensus can be perceived as clearer and less biased \cite{tessler2024ai}. In multi-party negotiation, AI moderation can raise fairness and efficiency perceptions \cite{kobiella2025ai}. \rev{However, delegating shared tasks can disrupt the mutual \emph{cooperative awareness} that groups use to coordinate actions and infer counterpart intent~\cite{dourish1992awareness}. This introduces a mismatch between computational optimization and the flexible, norm-sensitive realities of human group work~\cite{ackerman2000socio_technical}. We compare voluntary delegation against advisory and coaching modalities, evaluating how ceding control to an agent shapes cooperative dynamics and welfare outcomes in an interdependent game.}

\subsection{Adoption and Control in AI Assistance}
Despite potential benefits of AI adoption and delegation, takeup of such systems by human users often involves a calculated tradeoff between outcome quality and perceived control \cite{ryan2000self,owens2014control}.  
Human–AI teams may fail to achieve synergy even when AI improves accuracy \cite{green2019principles,vaccaro2024combinations}, in part because designs that enhance objective performance may increase cognitive effort or monitoring costs \cite{davis1989perceived,sweller1988cognitive,bucinca2020proxy,bucinca2021trust, qian2024take}.  
Users must also maintain accurate mental models of when AI is likely to err \cite{bansal2019beyond}; miscalibrated confidence or overestimation of personal ability \cite{yang2016,kruger1999unskilled} can lead to rejection of useful assistance. \textit{Algorithm aversion} further reduces adoption after observing small errors \cite{dietvorst2015algorithm,dietvorst2018overcoming}, though allowing human-in-the-loop intervention can restore perceived agency.

\rev{This agency-control tension is further shaped by the nature of the task itself and the organizational structures in which it is deployed. In knowledge acquisition and learning, higher levels of AI assistance can inadvertently depress human cognitive engagement, presenting an "AI Assistance Dilemma" where moderate, scaffolding support outperforms full automation~\cite{chen2025ai_assistance_dilemma}. Conversely, in strategic, mixed-motive environments like resource allocation, ceding complete decision initiative to a delegated agent may bypass human coordination inefficiencies, though human participants may still desire controllability and accountability over algorithmic actions. Realizing collective welfare gains therefore depends on understanding how these task dependencies and organizational dynamics shape the willingness to delegate decision initiative.}

In sum, these studies emphasize that adoption should not be treated as a proxy for effectiveness: users choose workflows that balance minimizing cognitive load and maximizing objective performance \cite{fugener2022cognitive,pathak2024ai}.  
Our work connects these insights to a strategic, multi-user environment: we examine how different human-AI interaction modalities shape adoption behavior, individual outcomes, and \rev{group-level externalities~\cite{grudin1988cscw_fail}} that arise when participants' decisions are mutually interdependent.

\section{Experiment}

We frame our empirical inquiry through three sets of research questions. Broadly, when access to AI is given to individuals in a multi-user context, who uses it, what do they gain, and what mechanism drives the welfare effects?

\paragraph{RQ1: Welfare effects.}
At the group level, does access and adoption of LLM assistance---in any of the three modalities (\textit{Advisor}, \textit{Coach}, \textit{Delegate})---increase \emph{individual}- and \emph{group}-level surplus, relative to a no-AI baseline?

\paragraph{RQ2: Adoption and preferences.} How do participants' stated preferences and usage decisions vary across the Advisor, Coach, and Delegate modalities, independent of their welfare effects?

\paragraph{RQ3: Mechanisms.} What mechanisms explain the differential welfare effects across modalities? Specifically, how does human intervention in Advisor and Coach modes alter the quality of AI-generated proposals, and to what extent does this filtering account for the Delegate advantage?  

\subsection{Game Setting}\label{sec:game_setting}

We explore the negotiation dynamics in the context of a chip bargaining game introduced by \citet{qian2025understanding}, where participants take turns exchanging chips of different colors, with randomly assigned private valuations, with the goal of maximizing their surplus. During each turn, one participant \textit{proposes} an offer (e.g. 7 red for 3 green chips), and the other two simultaneously and privately decide whether to \textit{accept} or \textit{decline}. If both accept, one is chosen randomly to clear the trade. The game ends after nine turns; participants leave with any surplus earned beyond the initial value of their chips. \rev{This stylized resource-exchange game models real-world collaborative tensions---cross-departmental scheduling, budget negotiations, and multi-party project coordination---where players must find Pareto-improving allocations under asymmetric information and conflicting interests.}

\begin{figure*}[htp]
    \centering
    \includegraphics[width=\linewidth]{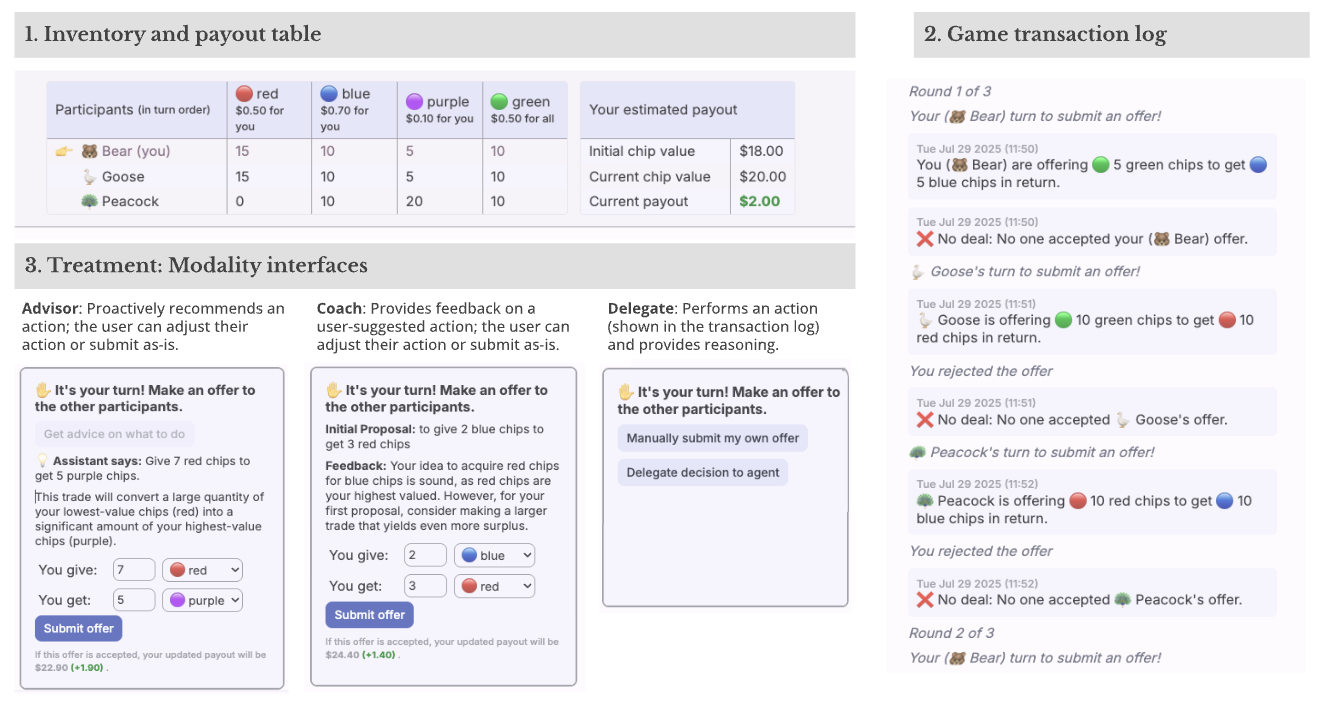}
    \caption{Relevant game components. Panels 1. and 2. show interface properties visible across all modalities. Panel 3 illustrates the proposal generation interface shown in each of the treatment modalities. More detailed figures of game interfaces are provided in Appendix~\ref{app:dl-interface}.}
    \label{fig:treatments}
\end{figure*}

This environment provides several properties that enable our investigation. First, by endowing participants with pre-defined preferences, it provides an objective ground truth for performance and outcome measurement, which is a requirement in empirical trading games exploring similar behaviors \cite{kahneman1990experimental, roth1991bargaining, bochet2024beyond}. Second, the combination of information asymmetry and restricted communication create a setting that necessitates strategic reasoning.\footnote{The complexity of this game prevents any dominant strategies that can be calculated ex-ante \cite{qian2025understanding}.} Finally, because players' outcomes are coupled through the channels detailed below, the multi-player, non-zero-sum design permits direct measurement of the group-level externalities that individual-task designs cannot isolate.

\paragraph{Strategic interdependence.} Most empirical evidence on human--AI collaboration, and in particular the bulk of the delegation and algorithm-aversion literature, comes from \emph{individual} decision-making tasks in which one user interacts with one AI system and the outcome depends only on that dyad \cite{agarwal2023combining, green2019principles, bansal2019beyond, dietvorst2015algorithm, vaccaro2024combinations}. Canonical domains---medical diagnosis, credit scoring, forecasting, deception detection---admit a single ground-truth target that the user and the assistant jointly estimate, and the assistance modality affects only the decision-maker's own payoff. Our game departs from this class in three specific ways:
\begin{itemize}
    \item \textbf{Payoff coupling.} A player can realize surplus only through trades that a counterparty agrees to execute; each player's final payoff is produced jointly with at least one other seat.
    \item \textbf{Opportunity-set externality.} Every consummated trade permanently alters the inventories of the two trading players, which shrinks or expands the set of feasible trades available on every subsequent turn---including the turns of the non-participating third player.
    \item \textbf{Informational externality.} Every proposal and every accept/decline decision reveals partial information about the actor's private valuations. All three players update beliefs from the public trade history, so one player's action shifts the best responses of the other two for the remainder of the game.
\end{itemize}
Because our design randomizes the assigned assistance modality at the group level (where all three players in a given session receive access to the same modality), it enables us to identify both the group-level Intent-to-Treat (ITT) effect of making a technology available and the individual-level \revnew{Voluntary non-compliance-adjusted estimate} of actual adoption under peer strategic spillovers.

\paragraph{Game-theoretical benchmark.}
Furthermore, the structure of the game allows computation of a Pareto-efficient benchmark for total surplus. \citet{qian2025understanding} measures the \textit{performance}, or surplus gain, as the surplus achieved by a group or individual divided by the maximum possible (Pareto-efficient) surplus for that specific game configuration, allowing for normalized comparison across different games and conditions. Throughout the paper, we refer to these measures as \textit{scaled group surplus} and \textit{scaled individual surplus}. 

\paragraph{Human Baseline.}
For our baseline, we reference the human-only performance established in \citet{qian2025understanding}. To ensure comparability, we verified that the recruitment criteria, payout rates (\$10 base + bonus), and interface mechanics were identical to the current study.
They found that human-only groups ($N=72$) achieve a mean scaled group surplus of 0.537 ($\pm$ 0.024). This suboptimal performance was attributed to human tendencies toward conservative trades and a ``fairness'' norm (e.g., 1-for-1 swaps), which systematically limited collective gains.

\subsection{Conditions}

We compare three treatment arms --- \textit{Access-to-Advisor}, \textit{Access-to-Coach}, and \textit{Access-to-Delegate}--- against two no-AI baselines from \citet{qian2025strategic}: the Human Baseline and a \textit{theoretical optimum} baseline (a linear-programming solution under full public valuations) that benchmarks optimized performance.

 The intervention is \textit{access} to a given treatment modality; in each game, each player can choose between taking action independently of AI assistance, or using the available AI intervention. All players within the game have access to the same intervention modality (e.g. a \textit{Delegate} game). A player can only interact with one modality in a given game and can choose whether to use assistance on each turn. \rev{We hid whether players used AI assistance from their counterparts. Players interacted under randomly-assigned, pseudonymous avatars (e.g. Bear, Bird), and the interface gave no indication whether a proposal or response was manual or AI-assisted. Counterparts observed only the proposed chip quantities and rationales, ensuring that responses reflected the objective quality of offers rather than biases toward algorithmic agents.}

\begin{itemize}
\item \textbf{Advisor:} The advisor agent proactively recommends an action, either an offer or an accept/reject response, and provides a rationale. The user can accept or revise this recommendation before submitting it.

\item \textbf{Coach:} The user first composes their intended action. The coach agent provides feedback on the user's plan, which the user can incorporate or disregard.
\item \textbf{Delegate:} The delegate agent autonomously generates and executes an action on the user's behalf. The user cannot veto or modify the AI's decision, but can view the agent's rationale.
\end{itemize}

Figure~\ref{fig:teaser} visualizes the experiment design. First, participants are introduced to the game rules and AI modalities, completing mandatory comprehension checks to ensure understanding. The instructions intentionally avoided \rev{revealing the AI's performance advantage} to prevent biasing towards delegation. The participants completed a pre-game survey and played three games in succession, each with access to a different AI assistance modality (Advisor, Coach, Delegate) in randomized order. \rev{To maintain game synchrony, we enforced a 60-second turn-timer for all proposal turns. The external Gemini API's average latency of 10.55 seconds consumed 18.3\% of this decision window. In the Advisor and Coach modes, this latency reduced the time available for users to review recommendations and draft offers, whereas the Delegate modality executed transactions directly.} Finally, the participant completed a post-game survey.

\subsection{LLM Agent Design and Implementation}

All assistance modalities were powered by an LLM-based agent, developed with two design goals in mind: i) its capabilities in the bargaining game must exceed human baselines, and ii) it needs to respond to the user with minimal latency. Our agent used the commercially-available \texttt{Gemini-2.5-Flash} API, with an output token limit of 8{,}192, thinking token budget of 2{,}048, and a throughput limit of 1{,}000 tokens/second to ensure responsive assistance. For robustness, model queries were implemented with two retries and a fallback mechanism that reverted users to manual override mode when necessary. Empirically, this implementation yielded a 99.71\% assistance success rate across 2,519 queries, with manual fallback triggered only seven times. 

To guarantee well-structured outputs, we applied constrained decoding to enforce JSON formatting (Appendix~\ref{app:Structured_offer}, Appendix~\ref{app:format_response}). 

While more powerful models like \texttt{Gemini 2.5-Pro} achieved slightly higher performance in simulations, their $>$40-second response latency was impractical for feedback during a real-time game. \texttt{Gemini-2.5-Flash} provided an average response time of 10.55 seconds. While this latency meets system-level usability guidelines for keeping a web user's attention active~\cite{nielsen1994usability, miller1968response}, it represents a severe temporal penalty inside a synchronous 60-second game turn. As discussed in Section~\ref{sec:limits}, this task-level temporal friction created severe cognitive pressure that artificially depressed compliance in Advisor and Coach modes by biasing users under time constraints toward rapid manual actions. Delegate mode, by contrast, bypassed this temporal friction entirely by executing transactions directly. We selected \texttt{Gemini-2.5-Flash} as the optimal engineering compromise between model capability and response speed. 
All three assistance modes were implemented as lightweight scaffolds on top of the same underlying agent, with supplemental reasoning text.\footnote{The full prompts for these agents are in Appendix~\ref{app:prompt}, and screenshots from the interfaces are in Figure~\ref{fig:treatments}.}

\paragraph{All-Agent Baseline.} We developed prompt scaffolding that \rev{outperforms the Human Baseline reported by} \citet{qian2025understanding} \rev{in the All-Agent Baseline}. In simulation, our agent achieved a scaled surplus of 0.595 ($\pm$ 0.024), significantly outperforming the human baseline (0.537 $\pm$ 0.024).

\subsection{Design and Analysis}

We utilize a within-participant design. Each participant played three separate games with access to a mode in randomized order, alongside two other players assigned to the same mode.

We collected the following performance measures:
\begin{itemize}
    \item \textbf{Individual/group surplus gain}: participant's or group's surplus change relative to the original chip values.
    \item \textbf{Proposal and decisions}: trading proposals and other players' responses (reject/accept).
    \item \textbf{AI takeup per turn}: whether the participant used AI assistance at certain turn.
\end{itemize}

We also collected self-reported subjective measures.
\paragraph{Pre-game survey.}
To inform RQ2, we employed a pre-game survey to capture user attributes.\footnote{A complete list of pre- and post- game survey questions and responses is provided in Appendix~\ref{app:survey_analysis}.} This survey captured three constructs on a 5-point Likert scale:

\begin{itemize}
    \item \textbf{Trust in AI}: Perceived ability, trustworthiness, insight, and helpfulness \cite{glikson2020human}. For analysis, we compute a composite pre-game trust score as the simple average of these four items.
    \item \textbf{Prior expertise}: Prior familiarity in games or tasks similar to the bargaining game.
    \item \textbf{Confidence}: Confidence in their ability to play the bargaining game well.
\end{itemize}

\paragraph{Post-game survey.}

Participants completed a survey after each of the three game rounds, and a final comparative survey. These surveys aimed to capture the following constructs:

\begin{itemize}
    \item \textbf{Satisfaction}: Satisfaction with final trading outcomes. 
    \item \textbf{Mental effort}: Cognitive load exhibited in the bargaining games.
    \item \textbf{Preference}: Choice of which AI modality they would prefer to use for future games.
\end{itemize}

Because assignment to an AI modality granted only \emph{access} to the assistant rather than enforcing its \emph{use}, compliance was voluntary and we report two complementary causal estimands. First, \textbf{Intent-to-Treat (ITT)} effects estimate the impact of being assigned access to a modality; we fit these using linear mixed-effects models with modality as a fixed effect and a random intercept for group, to account for the non-independence of the three participants within a negotiation. Second, at the individual level we report the \revnew{\textbf{Voluntary non-compliance-adjusted estimate}}, \revnew{calculated by dividing the individual-level ITT estimate by observed proposal compliance}. Full specifications are given in Section~\ref{sec:welfare-results}. Each family of ITT comparisons involves three simultaneous pairwise tests (Advisor, Coach, and Delegate each against the no-AI baseline) per level of analysis, which inflates the family-wise Type-I error rate if treated as independent; we therefore apply the Holm--Bonferroni correction~\cite{holm1979simple} within each family. We note that Holm--Bonferroni is conservative and may produce false negatives, particularly when tests are correlated~\cite{nakagawa2004farewell, chen2017general}.

\subsection{Procedure and Participants}
The game interface was implemented and deployed using Deliberate Lab \citep{Tsai_Deliberate_Lab_Open-Source_2025}, an open-source experimentation platform.\footnote{Additional game interface implementation details are provided in Appendix~\ref{app:dl-interface}.} 324 participants were recruited from the Prolific recruitment platform under an IRB-approved protocol, with no additional selection criteria \citep{prolific2024}. 

Our final sample includes $N=243$ participants over $81$ groups of three,\footnote{Because the design required full group participation across all three games, we excluded any group in which one or more members failed to complete the session. \rev{Excluded participants did not differ significantly from completers on pre-game trust or prior expertise ($p > .10$), indicating no systematic selection bias.}} with ~13 to 15 groups in each of the six unique orderings across the three treatments, involving $6,561$ trading decisions. Participants received a \$10.00 base payment, plus a performance-based bonus that averaged \$4.50, for approximately 56.4 minutes of their time. The bonus, calculated as the average individual surplus across the three games, was designed to align participant incentives with surplus maximization.

\section{Results}
\subsection{LLM Delegation leads to the highest gains (RQ1)}\label{sec:welfare-results}
Because adoption of AI assistance was voluntary within each condition, participants assigned to an AI modality could choose not to use it (non-compliance). We therefore present two complementary estimands: Intent-to-Treat (ITT) effects that capture the causal impact of being \emph{assigned access} to a modality, and a \revnew{Voluntary non-compliance-adjusted estimate} at the individual level.

\subsubsection{Model specification}

We employed Linear Mixed-Effects Models (LMM) to account for the nested structure of repeated measures within negotiation groups:
\begin{equation}
    Y_{ij} = \beta_0 + \beta_{1} \cdot \text{Condition}_{ij} + u_{j} + \epsilon_{ij}
    \label{eq:lmm}
\end{equation}
where $Y_{ij}$ is the scaled surplus for session $i$ in group $j$; $\beta_0$ is the intercept (Human Baseline); $\beta_{1}$ is the vector of fixed effects for the \emph{randomly assigned} AI modalities; $u_{j} \sim \mathcal{N}(0, \sigma_u^2)$ is a random intercept for group $j$; and $\epsilon_{ij} \sim \mathcal{N}(0, \sigma_\epsilon^2)$ is the residual error. Because Condition is randomly assigned, $\hat\beta_1$ identifies the ITT effect. We applied Holm-Bonferroni corrections across the three pairwise comparisons. \rev{To account for the strategic interdependence among players, standard errors in all individual-level regressions are clustered at the negotiation group level.}

At the individual level, 28--38\% of participants assigned to an AI condition never used AI for proposals. We report the \revnew{\textbf{Voluntary non-compliance-adjusted (VNCA) estimate}}, calculated as the individual-level ITT estimate divided by observed proposal compliance:
\begin{equation}
    \revnew{\widehat{\tau}_{\text{VNCA},m} = \frac{\widehat{\beta}_{\text{ITT},m}}{\widehat{c}_m}}
    \label{eq:vnca}
\end{equation}
\revnew{Here $\widehat{\beta}_{\text{ITT},m}$ is the individual-level ITT estimate for modality $m$, and $\widehat{c}_m$ is the share of participants who used AI for at least one proposal in that modality.} Observed compliance rates were 72.0\% (Advisor), 61.7\% (Coach), and 64.6\% (Delegate). \rev{Regressions of actual proposal usage on assignment yielded $F(1,242)=212.4$ for Advisor, $F=126.5$ for Coach, and $F=142.8$ for Delegate (all $p<0.001$). To account for the nested strategic dependencies among players in the same session, standard errors are clustered at the negotiation group level.}

The three estimands we report differ in both specification and level of aggregation:
\begin{itemize}
    \item \textbf{Group-level ITT.} Eq.~\ref{eq:lmm} fitted at the group level, with mean group surplus as the outcome. Identifies the effect of \emph{making a modality available} to a three-player group.
    \item \textbf{Individual-level ITT.} The same LMM in Eq.~\ref{eq:lmm}, fitted at the individual level (individual surplus as the outcome), retaining the group random intercept \rev{and clustering standard errors at the group level}. Identifies the within-group, per-participant effect of assigned access.
    \item \revnew{\textbf{Voluntary non-compliance-adjusted estimate.}} Eq.~\ref{eq:vnca} scales the individual-level ITT estimate by observed proposal compliance \rev{with group-clustered standard errors}.
\end{itemize}
The non-compliance adjustment has no group-level counterpart. Random assignment occurred at the group level: every member of a group is assigned to the same modality, so there is no gap between assignment and treatment at that level---a group assigned to Delegate \emph{is} a Delegate group, regardless of how many members actually use the AI. The group-level ITT from Eq.~\ref{eq:lmm} therefore directly identifies the policy-relevant causal effect of making a modality available.

\subsubsection{Results}\label{sec:welfare-combined}

Table~\ref{tab:welfare-combined} presents the combined welfare results, organized by the strength of statistical evidence.

\begin{table*}[ht]
\centering
\begin{tabular}{llrrrrl}
\toprule
\textbf{Level} & \textbf{Estimand} & \textbf{Modality} & \textbf{Coef.} & \textbf{SE} & \textbf{$p$} & \textbf{$p_{adj}$} \\
\midrule
\multicolumn{7}{l}{\emph{Suggestive trends (significant in uncorrected tests or approaching significance)}} \\
Group & ITT & Delegate & 0.084 & 0.040 & 0.033* & 0.100 \\
Individual & \revnew{\shortstack{VNCA estimate}} & Delegate & 0.043 & 0.023 & 0.061 & \rev{0.183} \\
Individual & ITT & Delegate & 0.028 & 0.015 & 0.067 & 0.200 \\
\midrule
\multicolumn{7}{l}{\emph{Non-significant}} \\
Group & ITT & Advisor & 0.006 & 0.040 & 0.888 & 1.000 \\
Group & ITT & Coach & 0.026 & 0.040 & 0.506 & 1.000 \\
Individual & \revnew{\shortstack{VNCA estimate}} & Advisor & 0.003 & 0.016 & 0.874 & \rev{1.000} \\
Individual & \revnew{\shortstack{VNCA estimate}} & Coach & 0.014 & 0.019 & 0.451 & \rev{0.902} \\
\bottomrule
\end{tabular}
\caption{Combined welfare results for RQ1. Standard errors are clustered at the negotiation group level. The top panel reports suggestive trends that are significant before multiple-testing corrections or approach conventional thresholds. Compliance rates: Advisor 72.0\%, Coach 61.7\%, Delegate 64.6\%.}
\label{tab:welfare-combined}
\end{table*}

\paragraph{Individual-level \revnew{Voluntary non-compliance-adjusted estimate} for Delegate (suggestive trend).}
\rev{Among participants in games with access to Delegation, the \revnew{VNCA estimate} for the Delegate modality is associated with an increased individual surplus of 0.043 scaled units. This estimate is roughly 1.5 times the corresponding uncorrected individual-level ITT estimate ($\hat\beta = 0.028$, $p = 0.067$). This suggestive estimate represents a compound effect of a participant's own delegation and the positive spillovers generated by their delegating peers, rather than an isolated individual utility gain.}

\paragraph{Supportive evidence: group-level and individual-level ITT}
At the group level, Delegate access produced a positive ITT effect ($\hat\beta = 0.084$, $p = .033$) that did not survive the Holm-Bonferroni correction ($p_{adj} = .10$). The individual-level ITT for Delegate showed a similar directional trend ($\hat\beta = 0.028$, $p = .067$, $p_{adj} = .20$). Neither the Advisor nor the Coach conditions showed any improvement over the baseline at either level.~\footnote{\rev{We evaluated chronological ordering to check for learning or fatigue; t-tests show no game-order effects (Appendix~\ref{app:ordering-effects}).}}

\rev{Although the multiple-comparison corrections render the individual \revnew{Voluntary non-compliance-adjusted estimate} marginally non-significant ($p_{\text{adj}} = 0.18$), the directional trend is consistent across all estimands: Delegate is the unique modality that shows positive coefficients at every level of analysis (group ITT, individual ITT, and the individual \revnew{Voluntary non-compliance-adjusted estimate}). The Advisor and Coach conditions produce null effects across all specifications.}

\begin{figure}[t]
    \centering
\includegraphics[width=1\linewidth]{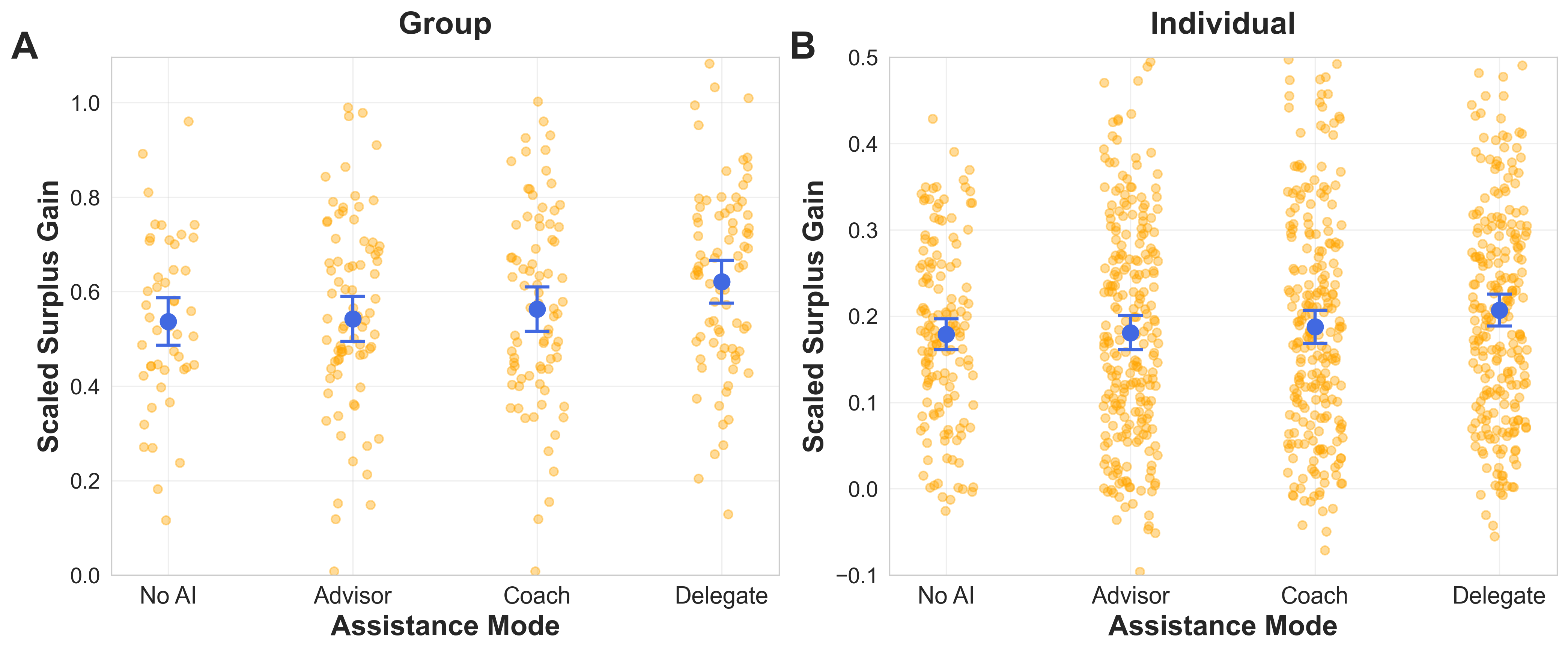}
    \caption{\textbf{\rev{Delegate access shows positive group-level trends and suggestive individual welfare gains; Advisor and Coach do not.}} \rev{(A) Group-level ITT: Delegate increases group surplus ($\hat\beta = 0.084$, $p = 0.033$, $p_{adj} = 0.10$); Advisor and Coach are indistinguishable from the baseline. (B) Individual-level ITT: a similar benefit for Delegate. Error bars represent 95\% confidence intervals clustered at the negotiation group level ($81$ active groups, $72$ baseline groups). Consistent trends support the ``human filter'' mechanism: retaining active control (Advisor/Coach) dilutes welfare benefits, while removing individual intervention (Delegate) translates model capability into realized surplus.}}
    \label{fig:surplus_comparison}
\end{figure}

\subsection{Users Prefer the Advisor Modality (RQ2)}

We analyzed participants' responses to the multiple-choice survey question: ``If you were to play again, which AI assistance mode would you prefer to use?''. Previous literature suggests that users would prefer higher-control modalities (Advisor and Coach) over the lower-control Delegate mode, despite the latter yielding the highest welfare (Section~\ref{sec:welfare-combined}).

\subsubsection{Results.}
A Chi-Square test revealed significant differences in user preferences across the three AI modalities ($\chi^2 = 45.03, p < .001$). To identify specific drivers of this preference, we conducted pairwise comparisons with Holm-Bonferroni corrections. The results provide \textbf{partial support} for RQ2, revealing a strong preference for the Advisor but not for the Coach as shown in Table~\ref{tab:prefrence}:
\begin{itemize}
    \item \textbf{Advisor vs. Delegate:} Consistent with RQ2, users significantly preferred the Advisor mode ($N=107$) over the Delegate mode ($N=47$) ($p_{adj} < .001$).
    \item \textbf{Advisor vs. Coach:} Users also overwhelmingly preferred the Advisor over the Coach ($N=37$) ($p_{adj} < .001$).
    \item \textbf{Delegate vs. Coach:} There was no significant difference in preference between the Delegate and the Coach ($p_{adj} = .326$).
\end{itemize}

\begin{table}[!htbp]
\centering
\setlength{\tabcolsep}{6pt}
\begin{tabular}{lcc}
\toprule
Preferred Mode & Number of Participants ($N$) & Share (\%) \\
\midrule
\quad Advisor & 107 & 44.0 \\
\quad Coach   & 37  & 15.2 \\
\quad Delegate & 47 & 19.3 \\
\quad None     & 52 & 21.4 \\
\bottomrule
\end{tabular}
\caption{Preference differences across agent types: users dominantly prefer the Advisor modality.}
\label{tab:prefrence}
\end{table}

\subsubsection{Rationales for preferences}\label{sec:rationale}
We furthermore conducted a semantic thematic analysis of the open-ended rationale texts following the standard Braun \& Clarke framework~\cite{braun2006thematic}. We identified four recurring themes for preference: (i) \textit{trust and control}, (ii) \textit{ease of use and cognitive offloading}, (iii) \textit{effectiveness and performance}, and (iv) \textit{other}.\footnote{Thematic analysis methodology provided in Appendix~\ref{app:classification}.}

\textbf{Advisor-preferrers} frequently credited the AI's effectiveness and performance as the reason for their selection; many also referred to trust and control and ease of use. From P92:

\begin{quote}
\textit{``I love the advisor. It helps when you get into the weeds of the game when the strategies become less obvious. I also like that I still have full control.''}
\end{quote}

Participants who preferred none of the modalities (``\textbf{autonomy-seekers}'') also cited trust and control, but as grounds for disengagement. From P96:
\begin{quote}
\textit{``I don't trust AI bots; I feel I can make better decisions on my own.''}
\end{quote}

\textbf{Delegate-preferrers} valued ease of use and cognitive offloading. From P47 and P99:
\begin{quote}
\textit{``Delegate helped with ease of decision making and made it easiest for me.''}
\end{quote}
\begin{quote}
\textit{``I prefer that someone else make the decision.''}
\end{quote}

\textbf{Coach-preferrers} highlighted effectiveness and performance as the main rationale for their selection. From P266:
\begin{quote}
\textit{``Coach helped me see things that I didn't see myself like a real coach.''}
\end{quote}

\subsection{Human Intervention as a Mechanism for Welfare Differences (RQ3)}\label{sec:mechanism}

RQ1 establishes that Delegate access produces welfare gains while Advisor and Coach do not; RQ2 shows that users nonetheless prefer the Advisor. All three modalities are powered by the same underlying LLM, so the performance gap must arise from differences in how the AI's output reaches the market. In this section, we decompose the treatment effect by examining (i)~the quality of AI-generated versus manual proposals, (ii)~the degree to which human intervention filters AI output, and (iii)~the resulting trade composition across conditions.

\subsubsection{AI-generated proposals create more surplus}

We classify each accepted trade by whether the proposer used AI assistance, and compute the joint surplus created (sum of sender and recipient payout changes). Table~\ref{tab:mechanism-quality} reports the results. Across all three modalities, accepted trades originating from AI-assisted proposals generate higher joint surplus than manual proposals. The gap is largest in the Advisor condition ($+0.52$). Because non-AI offers produce comparable surplus across conditions ($\sim$2.7--3.1), the modality-level performance gap is driven primarily by how much of the AI's proposal quality reaches the final offer.

\begin{table}[H]
\centering
\begin{tabular}{lccccc}
\toprule
& \multicolumn{2}{c}{\textbf{AI-Assisted Offers}} & \multicolumn{2}{c}{\textbf{Manual Offers}} & \\
\cmidrule(lr){2-3}\cmidrule(lr){4-5}
\textbf{Mode} & \textbf{Joint Surplus} & $N$ & \textbf{Joint Surplus} & $N$ & \textbf{$\Delta$} \\
\midrule
Advisor  & 3.201 & 158 & 2.678 & 157 & +0.523\rev{$^{**}$} \\
Coach    & 2.900 & 157 & 2.746 & 200 & +0.154 \\
Delegate & 3.290 & 160 & 3.110 & 186 & +0.180 \\
\bottomrule
\end{tabular}
\caption{Mean joint surplus per accepted trade, stratified by whether the proposer used AI assistance. Significance markers represent pairwise t-tests with group-clustered standard errors ($^{**}p<0.01$). The difference is statistically significant only in the Advisor condition ($p < 0.01$). \rev{The total number of accepted trades across Advisor ($N=324$), Coach ($N=372$), and Delegate ($N=352$) games differs slightly from the sum of trade categories due to a minor percentage of turns (Advisor: 9; Coach: 15; Delegate: 6) in which model API failures or network dropouts temporarily triggered an automatic manual fallback, which have been excluded from this categorization.}}
\label{tab:mechanism-quality}
\end{table}

\paragraph{Evidence of Market-Wide Positive Spillovers \& Distributional Equity.} \rev{A key finding in Table~\ref{tab:mechanism-quality} is the presence of significant spillovers. Manual proposals in the Advisor ($2.678$) and Coach ($2.746$) conditions generate surplus comparable to the baseline, yet manual proposals in Delegate games yield substantially higher surplus ($3.110$, $p < 0.01$, post-hoc t-test). The presence of Delegate agents thus lifts the quality of all accepted trades, including purely manual ones.}

\rev{This spillover effect resolves a key empirical paradox: why does the Advisor modality fail to generate positive spillovers or group-level welfare gains, even though its accepted AI-assisted offers are of virtually identical quality ($3.201$ joint surplus) and slightly higher volume ($N=158$) than those in the Delegate modality ($3.290$ joint surplus, $N=160$)? Table~\ref{tab:mechanism-quality} proves that this discrepancy is entirely driven by the quality of manual proposals. In Delegate games, unassisted human-to-human trades achieved a significantly higher surplus ($3.110$) than in Advisor ($2.678$) or Coach ($2.746$) games ($p < 0.01$). By taking over proposals entirely on delegation turns, the Delegate agent cleared conservative baseline offers without human intervention. In the Advisor and Coach modes, by contrast, the human filter actively diluted the quality of offers: users modified or ignored optimal recommendations, pulling the shared offer pool back toward conservative human baseline distributions and preventing the emergence of these positive spillovers.}

\rev{This spillover carries important distributional implications. A central concern with deploying AI agents in strategic settings is transactional exploitation---an AI delegate might maximize aggregate surplus through asymmetric, predatory splits that extract value from unassisted counterparts. Although joint surplus alone cannot rule out payoff asymmetry, our data provides an empirical example against this concern.}

\paragraph{\rev{Individual-Level Regression of Peer Spillovers.}} \rev{To formally test whether the Delegate modality generates individual-level spillovers for non-users, we estimated a Linear Mixed-Effects Model comparing the scaled individual surplus of non-adopters in AI conditions against the Human Baseline. We define a non-adopter as a participant who did not use the AI assistant for any proposals in that specific game. The model is specified as:}
\begin{equation}
    Y_{ij} = \beta_0 + \beta_{1} \cdot \text{Delegate\_NonUser}_{ij} + \beta_{2} \cdot \text{Advisor\_NonUser}_{ij} + \beta_{3} \cdot \text{Coach\_NonUser}_{ij} + u_{j} + \epsilon_{ij}
    \label{eq:spillover-lmm}
\end{equation}
\rev{where $Y_{ij}$ is the individual surplus, $\beta_0$ represents the mean of the Human Baseline ($0.179$), and coefficients $\beta_{1}, \beta_{2}, \beta_{3}$ capture the marginal effect of being a non-adopter in the Delegate, Advisor, and Coach conditions, respectively, with a random intercept $u_j$ for group and standard errors clustered at the group level.}

\rev{As reported in Table~\ref{tab:spillover_results}, non-adopters in Delegate games achieved a mean scaled individual surplus of $0.218$, a $21.6\%$ increase over the baseline. The LMM regression identifies a positive coefficient for Delegate non-adopters ($\beta = 0.039, SE = 0.020$, uncorrected $p = 0.054$) that approached statistical significance in direct comparison, though it did not survive multiple-comparison correction ($p_{\text{adj}} = 0.16$). Non-adopters in Advisor and Coach games showed no improvement over the baseline. Notably, Delegate non-adopters achieved a higher average surplus than active AI adopters within the same games ($0.218$ vs.\ $0.201$).}

\rev{We caution, however, against a purely causal interpretation of this non-adopter coefficient, which is subject to selection bias. Because adoption was voluntary, the decision to reject AI assistance is endogenous. Pre-game survey data (Section~\ref{sec:trust}) shows that non-adopters (autonomy-seekers) reported distinct profiles, such as lower post-game mental effort. This suggests that the Delegate non-adopter coefficient conflates a genuine market spillover with a selection effect, where high-ability bargainers self-selected into manual play when assigned to Delegate sessions. Indeed, if the spillover were purely driven by the presence of high-quality AI proposals in the market, Advisor games---which generated a comparable volume ($N=158$) of high-surplus accepted AI trades---should show similar spillovers. Yet Advisor non-adopters show a negative, non-significant coefficient ($\beta = -0.017$, $p = 0.443$). This divergence strongly suggests that the observed Delegate non-adopter benefit is at least partially driven by endogenous selection into manual play rather than system externalities alone. Future work employing randomized mandatory-delegation blocks is required to isolate these causal channels.}

\begin{table}[h]
\centering
\begin{tabular}{lcccc}
\toprule
\textbf{Non-Adopter Condition} & \textbf{Coef. ($\beta$)} & \textbf{Std. Error} & \textbf{z-value} & \textbf{P-value} \\ 
\midrule
Intercept (Human Baseline) & 0.179 & 0.012 & 14.635 & 0.000 \\
Delegate Non-Adopter (vs Baseline) & \rev{0.039} & \rev{0.020} & \rev{1.926} & \rev{0.054} \\
Advisor Non-Adopter (vs Baseline) & \rev{-0.017} & \rev{0.022} & \rev{-0.768} & \rev{0.443} \\
Coach Non-Adopter (vs Baseline) & \rev{0.012} & \rev{0.020} & \rev{0.627} & \rev{0.530} \\ 
\bottomrule
\end{tabular}
\caption{\rev{Linear Mixed-Effects Model results comparing individual surplus of non-adopters in AI conditions against the Human Baseline ($N=243$ active individuals nested in $81$ groups; $N=216$ baseline individuals nested in $72$ groups). Standard errors are clustered at the negotiation group level. The listed p-values are uncorrected; the Delegate coefficient does not survive Holm-Bonferroni correction ($p_{\text{adj}} = 0.16$).}}
\label{tab:spillover_results}
\end{table}

\subsubsection{Human intervention filters AI quality}\label{sec:filtering}

Even when participants requested AI assistance, adoption of the AI's suggestions was far from automatic. The human filter between the AI's recommendation and the submitted action operates through three distinct channels: selective adoption, modification of suggestions, and ignoring feedback.

\paragraph{Selective adoption.}
Participants used AI for offer generation 54.9\% of the time in Advisor games, 43.4\% in Coach, and 48.5\% in Delegate. AI usage was significantly higher for offer generation than for offer responses across all modalities ($p < .001$; Table~\ref{tab:overall_offer_response}), consistent with participants perceiving proposals as the more consequential decision. In non-AI turns, proposals resembled the human baseline (mean trade size $\sim$10.3 chips versus $\sim$12--13 for AI-assisted offers). Higher pre-game trust in AI predicted greater AI usage across all modes ($r \sim .25$, $p < .01$).

\begin{table}[h]
\centering
\begin{tabular}{lrrrr}
\hline
\textbf{Mode} & \textbf{Offer Freq.} & \textbf{Res.\ Freq.} & \textbf{t-stat} & \textbf{p-value} \\
\hline
Coach    & 0.434 & 0.273 &  6.617 & <0.001 \\
Advisor  & 0.549 & 0.285 & 11.008 & <0.001 \\
Delegate & 0.485 & 0.264 &  9.184 & <0.001 \\
\hline
\end{tabular}
\caption{AI takeup rates for offer generation and offer responses at the participant level ($N=243$ individuals nested in $81$ groups). The t-statistics and p-values represent paired t-tests at the participant level comparing offer-side vs. response-side takeup rates. AI usage is significantly higher for generating offers than for responding across all modalities ($p < .001$).}
\label{tab:overall_offer_response}
\end{table}

\paragraph{Modification of AI suggestions (Advisor).}
Among Advisor users who requested AI assistance for proposals, 70.6\% submitted the AI's recommended offer without modification; 29.4\% modified it before submission. When modifying, participants tended toward conservatism: 49.1\% reduced the trade size, 37.1\% increased it (net change: $-0.58$ chips). For offer \emph{responses}, the acceptance rate of AI-recommended actions was lower still, indicating that human override is pervasive on both the proposal and response sides.

\paragraph{Ignoring coaching feedback (Coach).}
In Coach mode, participants drafted their own proposal first, then received AI feedback. Only 30.5\% of users changed their offer after receiving coaching. On the response side, the filter was even stronger: users retained their initial accept/reject decision 96\% of the time, even when the AI recommended the opposite action. The coaching intervention had minimal influence on final decisions.

\begin{figure}[htp]
    \centering

    \begin{minipage}{0.65\linewidth}
        \centering
        \includegraphics[width=\linewidth]{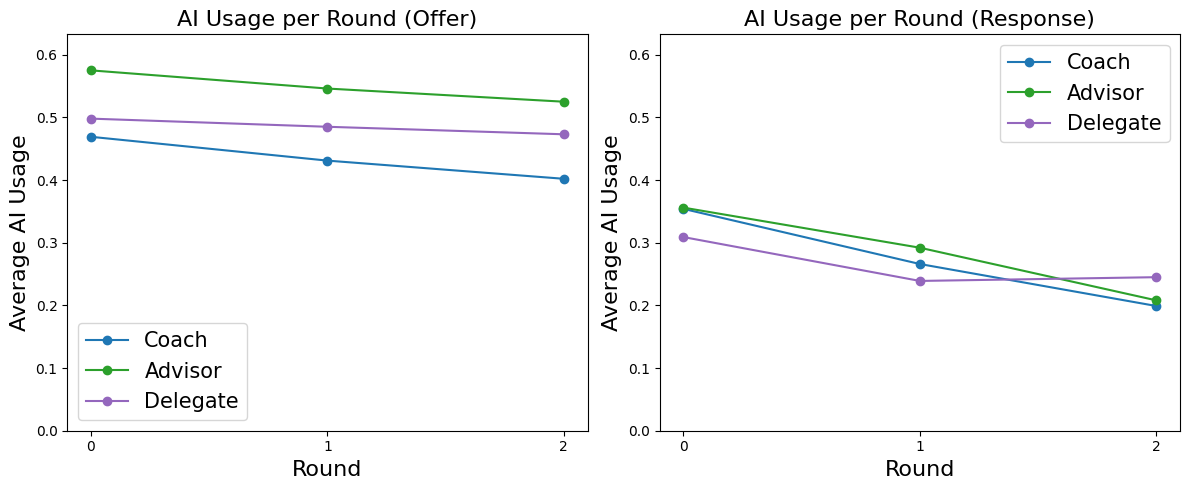}
    \end{minipage}%
    \hfill

    \begin{minipage}{0.5\linewidth}
        \centering
        \begin{tabular}{lccc}
        \toprule
        \textbf{Mode} & \textbf{Offer} & \textbf{Response} \\
        \midrule
        All   & $-0.077(0.114)$ &     $-0.407^{***}$ (0.000)\\
        \midrule
        Coach  & $-0.053^{*}$ (0.049) & $-0.262^{***}$ (0.000) \\
        Advisor & $-0.034$ (0.148)     & $-0.248^{***}$ (0.000) \\
        Delegate & $-0.019$ (0.446)     & $-0.122^{*}$ (0.019) \\
        \bottomrule
        \end{tabular}
    \end{minipage}

    \caption{Top: Frequency of assistance usage by negotiation round. Bottom: Coefficient of regression predicting AI usage as a function of the negotiation round. Significance levels: $^{***}p<0.001$, $^{**}p<0.01$, $^{*}p<0.05$.}
    \label{fig:aiusage_regression}
\end{figure}

\paragraph{Temporal dynamics.}
As shown in Figure~\ref{fig:aiusage_regression}, AI usage for offer generation remained relatively steady across the three rounds of play, whereas usage for responses declined significantly in all modalities (coef.\ $= -0.407$, $p<0.001$). This divergence suggests that participants perceived ongoing value in AI-assisted proposal generation but quickly learned to rely on their own judgment for accept/reject decisions---a pattern consistent with the lower complexity and faster feedback cycle of binary response decisions.

\subsubsection{Trade composition across modalities}

Table~\ref{tab:mechanism-composition} summarizes the resulting trade-level outcomes. Delegate games produce the highest joint surplus per accepted trade (3.200). The Delegate advantage does not stem from a higher acceptance rate---the Coach has the highest (0.510)---but from the higher quality of each accepted trade, which is preserved because the AI's proposals reach the market without human modification.

\begin{table}[H]
\centering
\begin{tabular}{lcccc}
\toprule
\textbf{Mode} & \textbf{Mean Trade Size} & \textbf{Accept Rate} & \textbf{Joint Surplus/Trade} & $N_{\text{trades}}$ \\
\midrule
Advisor  & 11.74 & 0.444 & 2.937 & 324 \\
Coach    & 10.16 & 0.510 & 2.824 & 372 \\
Delegate & 11.19 & 0.483 & 3.200 & 352 \\
\bottomrule
\end{tabular}
\caption{Trade-level characteristics by modality.}
\label{tab:mechanism-composition}
\end{table}

\subsubsection{Summary}

Human intervention acts as a relational preservation filter on AI output, operating at every stage of the decision process. On the \emph{adoption} margin, participants bypass the AI entirely in 45--57\% of turns. On the \emph{modification} margin, Advisor users make proposals more conservative and Coach users largely ignore feedback. In socio-technical terms, these interventions align the AI's optimizing proposals with human social expectations, trading off raw material surplus for relational alignment in the Advisor and Coach conditions. The Delegate advantage arises not from a different AI capability, but from the absence of this relational mediation, allowing the AI's direct material optimization to reach the market.

\subsection{Explorative Analysis}

\subsubsection{Who adopts AI? The role of trust and cognitive load.}\label{sec:trust}

Table~\ref{tab:prior_confidence} reports pre-game and post-game self-reported measures by participants' stated modality preference. Two patterns stand out. First, participants who preferred \emph{any} AI mode reported 25\% higher mental effort than autonomy-seekers ($3.39$ vs.\ $2.71$, representing a 20\% reduction in effort for autonomy-seekers; $p < .01$), and higher mental effort was associated with lower surplus across all games (OLS coef $= -0.012$, $p = 0.023$). This suggests that participants who found the game more cognitively demanding were both more likely to seek AI assistance and less likely to perform well when acting independently---a complementarity that should, in principle, favor delegation. Second, autonomy-seekers reported the lowest pre-game confidence ($3.63$ vs.\ $4.00$--$4.25$ for AI-preferrers), suggesting that their rejection of AI stems from general skepticism rather than high self-assessed ability.

\begin{table}[ht]
\centering
\begin{tabular}{lccc}
\toprule
\textbf{Preferred Mode}
& \textbf{Confidence }
& \textbf{Prior Experience }
& \textbf{Mental Effort } \\
\midrule
Coach    & 4.00 $\pm$ 0.25 & 2.26 $\pm$ 0.31 & 3.46 $\pm$ 0.16 \\
Advisor  & 4.19 $\pm$ 0.16 & 2.34 $\pm$ 0.21 & 3.31 $\pm$ 0.10 \\
Delegate & 4.25 $\pm$ 0.23 & 2.42 $\pm$ 0.34 & 3.50 $\pm$ 0.17 \\
None     & 3.63 $\pm$ 0.30 & 2.25 $\pm$ 0.33 & 2.71 $\pm$ 0.18 \\
\bottomrule
\end{tabular}
\caption{Pre-game self-reported confidence and prior experience, and post-game reported mental effort (i.e., how hard participants found the game), by preferred AI mode.}
\label{tab:prior_confidence}
\end{table}

\section{Discussion}

Aggregate performance metrics can obscure how groups achieve outcomes and who benefits. We analyze our results through three core themes: how voluntary AI delegation represents a social dilemma of adoption; how the ``human filter'' acts as a socio-technical translator between raw efficiency and social norms; and how task structure and organizational hierarchy shape the optimal allocation of agency.

\subsection{Socio-Technical Dynamics and the Inverted Collaboration Paradox}

\rev{In collaborative systems research, Grudin's Paradox posits that technology fails to adopt because the individuals who bear the cost of system entry (e.g., manual data entry) receive zero individual utility, while others reap the benefits~\cite{grudin1988cscw_fail}. Our findings with the Delegate modality present a distinct, structural variation of this pattern. Because Delegate adopters receive suggestive, individual-level welfare gains ($\hat\beta = 0.043$, Table 1), delegation is materially a win-win scenario rather than a classic utility misalignment. The barrier to adoption is not structural (a lack of individual utility), but psychological: the delegating participant cedes agency and bears the psychological cost of relinquishing decision control, yet their adoption generates positive group externalities that benefit counterparts.}

\rev{While the group-level ITT trend is suggestive ($\hat\beta = 0.084$, $p = 0.033$, $p_{\text{adj}} = 0.10$), our trade-level decomposition in Section~\ref{sec:mechanism} reveals a powerful market-wide spillover: manual proposals in Delegate-access games yield higher joint surplus ($3.110$) than those in Advisor ($2.678$) or Coach ($2.746$) games ($p < 0.01$). The Delegate acts as a local ``market maker,'' introducing higher-quality, non-conservative proposals into the shared offer pool, lifting the welfare floor for adopters and non-adopters alike. Crucially, this spillover provides an empirical example against the central ethical concern of algorithmic exploitation in multi-party systems: that an optimization agent might maximize joint surplus by extracting value from unassisted counterparts through highly asymmetric splits. In our study, unassisted counterparts (non-adopters) in Delegate games actually achieved a higher average individual surplus ($0.218$) than the active AI adopters who utilized the Delegate ($0.201$). This empirical fact provides an instance of transactional equity: rather than acting as a predatory tool that exploits counterparts, the Delegate agent upgrades the shared decision space, permitting counterparts to capture equal or even greater welfare gains. Structurally, this creates a social dilemma of adoption: individual-level algorithm aversion and a desire for decision agency prevent the group from reaching a system-level, Pareto-efficient cooperative equilibrium. Designing collaborative systems in these settings is therefore a mechanism design problem: we must design interfaces and interaction rules that reduce these psychological ceding costs to unlock group-level welfare outcomes~\cite{grudin1988cscw_fail}.}

\rev{Why does delegation outperform advisory and coaching interfaces when using the identical underlying model? In the Advisor and Coach conditions, human intervention acts as a human filter: users bypass the AI, edit proposals toward conservative allocations, or ignore coaching feedback. In mixed-motive bargaining, the conservative, symmetric proposals that humans favor (such as 1-for-1 chip swaps) are not "inefficient failures" to maximize utility; they are vital signals of relational labor that establish trustworthiness, index reciprocity, and build long-term social capital. By optimization-forcing the proposals, the Delegate modality strips this relational buffer. While ceding control to the Delegate maximizes short-term economic efficiency, it does so by suppressing the social feedback loops that prevent cooperative breakdown over longer horizons. Furthermore, in real organizational hierarchies, retaining controllability is a rational professional action to take; delegating hig-stakes transactions in the real world at the cost of personal visibility risks a loss of accountability for un-audited algorithmic errors. The Advisor's modification window can therefore serve as a safe buffer; the human filter bridges Ackerman's fundamental mismatch between rigid computational optimization and the flexible, accountability-driven realities of social coordination~\cite{ackerman2000socio_technical}. Realizing collective welfare in collaborative systems depends on designing interfaces that honor this relational labor rather than seeking to bypass it entirely.}

\subsection{The Psychological and Structural Limits of Agency Allocation}

A recurring pattern across studies of multi-party bargaining is the tension between human agency and systemic efficiency. Prior work documents that human traders adhere to fairness norms that limit total surplus but may serve important social functions---maintaining cooperation, signaling trustworthiness, and sustaining norms of reciprocity~\cite{qian2025strategic}. In the present study, participants strongly prefer the Advisor modality (44\% of participants) over the Delegate (19\%), even though the Delegate produces the highest welfare gains. Qualitative rationales confirm that Advisor-preferrers value the sense of remaining ``in control,'' while Delegate-preferrers emphasize cognitive offloading.

\rev{This control premium can be interpreted post-hoc through Self-Determination Theory (SDT)~\cite{ryan2000self}, which posits that ceding complete authority to a Delegate agent may diminish a user's sense of autonomy and ownership. In collaborative tasks, the psychological reward of a successful outcome is tightly coupled with active decision-making. By automating proposals entirely, the Delegate modality reduces the user to a passive observer, potentially severing this reward. In contrast, the Advisor interface preserves the user’s agency, allowing them to derive satisfaction from exercising active judgment---or actively overriding the algorithm---even when doing so reduces objective material payoffs. While our survey did not formally measure psychological needs or self-efficacy scales, our open-ended qualitative data (Section~\ref{sec:rationale}) provides descriptive support for this interpretation: Advisor-preferrers overwhelmingly cited the value of retaining "full control" as their primary motive. Prior research shows that AI-to-human delegation (where an AI manager assigns tasks to a human) can boost human self-efficacy by aligning tasks with skills~\cite{hemmer2023human}. Our suggestive findings point to a potential psychological inversion in human-to-AI delegation: when the human cedes authority entirely, the loss of decision autonomy may act as a severe psychological barrier that outweighs objective economic gains, representing a critical avenue for future empirical measurement.}

This preference--performance misalignment generalizes earlier findings from individual decision-support tasks~\cite{bansal2019beyond, bucinca2021trust} to a multi-party strategic domain. The ``cognitive miser'' hypothesis~\cite{bucinca2021trust, owens2014control} would predict that users should favor the modality requiring the least effort---the Delegate. Instead, users accept the higher cognitive load of the Advisor to retain a sense of agency, suggesting that the control premium in strategic contexts outweighs the appeal of cognitive offloading. 

An alternative explanation is capability uncertainty: our instructions deliberately avoided revealing the AI's \rev{strong} performance, so participants entered with uncalibrated beliefs about the agent's competence. Pre-game trust indeed predicts higher takeup across all modes (Section~\ref{sec:trust}). Interfaces that surface model confidence, show counterfactual outcomes, or provide a veto window for delegated actions may narrow the gap between objective gains and realized adoption by explicitly calibrating user reliance and reducing algorithm aversion~\cite{bansal2021does, vaccaro2024combinations, lee2023counterfactual, vasconcelos2023explanations_reduce, green2019principles_limits}.

Of these candidate mechanisms, algorithm aversion~\cite{dietvorst2015algorithm} provides the most parsimonious account of the observed pattern. The Advisor's dominance in stated preference (44\%) maps directly onto the prediction of \citet{dietvorst2018overcoming}: users accept algorithmic assistance when they can modify its output, even if their modifications are welfare-reducing. The qualitative rationales reinforce this interpretation---Advisor-preferrers cite retained control (``I still have full control''), while autonomy-seekers reject the AI outright (``I don't trust AI bots'')---both of which are hallmarks of algorithm aversion rather than cognitive offloading or rational capability assessment. 

The cognitive miser hypothesis would predict preference for the Delegate, since it requires the least effort; capability uncertainty could in principle be resolved by revealing the AI's performance. Neither alternative accounts for why users actively prefer a modality that lets them override a superior system. Our results therefore extend the algorithm aversion literature from individual prediction tasks~\cite{dietvorst2015algorithm, dietvorst2018overcoming} to multi-party strategic settings, where the welfare cost of aversion is amplified: when one participant's refusal to delegate reduces not only her own surplus but also the quality of offers available to other players, algorithm aversion generates negative externalities that compound across the group.

\rev{Furthermore, the optimal allocation of agency depends on task structure. In knowledge acquisition, where human reasoning is the primary source of value, more AI assistance can reduce cognitive engagement, meaning users perform best with moderate assistance that preserves active reasoning (the ``AI Assistance Dilemma'')~\cite{chen2025ai_assistance_dilemma}. In strategic resource allocation, however, human reasoning introduces systematic departures from surplus-maximizing behavior by prioritizing social equity over Pareto efficiency. Here, full delegation may structurally outperform intermediate scaffolding. Designers must therefore assess where in the cooperative pipeline human judgment adds or subtracts value, mapping these choices onto systematic frameworks of task delegability~\cite{lubars2019ask_not}.}

\rev{Finally, deploying autonomous delegates in real-world scenarios can introduce complex hierarchical dynamics. Delegating high-stakes negotiations to an AI requires ceding professional visibility and risking accountability for un-audited mistakes. Autonomous delegation is blocked in practice by an accountability vacuum: if a delegated agent executes a transaction resulting in organizational loss, the blame vectors between the human principal, the algorithm, and the system developers remain highly ambiguous. Real-world adoption will be dictated not merely by system usability, but by how accountability structures shape the risk-acceptability of ceding agency.}

\subsection{Design Implications for Collaborative AI Systems}

Our findings suggest four design paths for multi-party AI assistance:

\begin{enumerate}
    \item \rev{\textbf{Delegation with veto windows and progressive trust.} The Delegate's welfare advantage stems from removing the human filter, but full delegation raises concerns about accountability, user autonomy, and cooperative awareness~\cite{amershi2019guidelines}. A ``veto window'' design---where the AI acts autonomously but the user can override within a review period---may preserve the welfare benefit while restoring perceived control. For example, a ``draft-and-hold'' queue could display the proposed trade and its rationale, releasing it autonomously unless vetoed. Designers must remain cognizant of cognitive monitoring costs: if the user feels compelled to audit every action, the delegation benefit is diluted.}

    \item \rev{\textbf{Progressive confidence disclosure.} Capability uncertainty drives non-adoption. Interfaces that progressively reveal the AI's track record (e.g., displaying performance summaries after an initial unbiased exposure period) may reduce algorithm aversion by calibrating user expectations. Surfacing dynamic credibility indicators can alter user reliance in cooperative tasks~\cite{lu2022effects_credibility}.}

    \item \rev{\textbf{Evaluating modalities at the group level.} Individual preference data alone can mislead system designers: the modality users prefer (Advisor) is not the modality that maximizes group welfare (Delegate). Evaluation frameworks should incorporate group-level telemetry---such as negotiation velocity, allocative equity, and Pareto efficiency gains---alongside individual satisfaction metrics~\cite{grudin1988cscw_fail}.}

    \item \rev{\textbf{Hybrid and adaptive modalities.} Systems could adaptively shift between modalities---defaulting to coaching on early turns (to build trust, calibrate mental models, and support cooperative awareness) and offering delegation on later turns (to capture efficiency gains). Designers must implement clear interface signaling during mid-session transitions to prevent lapses in cooperative awareness as users shift between active planning and passive monitoring.}
\end{enumerate}

\section{Limitations}

\subsection{Task Design and Ecological Realism}

\rev{We employ a stylized three-player chip-trading game to study agentic assistance under strategic interdependence. The game isolates key bargaining tensions---private information, mixed-motive incentives, and joint externalities---while permitting precise control over payoffs and counterfactuals~\cite{ficici2008coloredtrails, pfeffer2007reasoning}. However, this abstraction omits several real-world cooperative dynamics: unconstrained natural language communication, long-term relationships, and domain-specific corporate norms.} \rev{First, our study evaluates outcomes from single-session games with at most nine turns per player. This horizon is too brief to capture how trust, reliance, and strategy evolve over time; trust in automation is dynamic and history-dependent, shaped by accumulated successes and failures~\cite{glikson2020human}. Second, our scaled surplus metric captures short-horizon economic efficiency. In real-world collaborative work where sustained cooperation, trust-building, and perceived fairness carry long-term value, the human filter---which we characterize as relational preservation---may serve a vital welfare-preserving function by maintaining social norms. Our finding that delegation outperforms human-mediated modalities is bounded by this stylized task; extending it to environments where relational considerations dominate requires further empirical work.}

\subsection{Methodological and Econometric Constraints}

\rev{Our primary causal findings are identified by the counterbalanced within-subject variation across the three active treatment arms in our active sample ($N=243$). However, our comparisons to the all-human baseline rely on a historical control group ($N=216$) collected under identical platform and incentive settings in \citet{qian2025strategic}. While we verified that recruitment criteria, payment rates, and mechanics were identical, historical controls are subject to temporal drift, minor cohort shifts, or platform changes over time. Absolute cross-study comparisons (such as Table~\ref{tab:welfare-combined}) should therefore be interpreted with caution.}

\rev{Furthermore, the strategic interdependence of the game introduces a SUTVA violation}\revnew{~\cite{angrist1996identification}}\rev{: one player's treatment (Delegate access) alters the opportunity set of others through spillovers. Consequently, our }\revnew{Voluntary non-compliance-adjusted estimate}\rev{ ($\hat\beta = 0.043$) conflates the direct effect of own delegation with the indirect effects of peer delegation. While this SUTVA violation represents the exact collaborative externalities we seek to evaluate, future work using network-based designs or spillover-robust methods is required to mathematically separate these causal channels.}

\rev{Finally, our within-subject design counterbalanced modality assignments chronologically to control for position effects (first, second, third game), showing no significant position-based learning or fatigue ($p > 0.05$, Appendix~\ref{app:ordering-effects}). However, this does not fully rule out sequence-specific carryover effects. In mixed-motive group negotiations, a participant who experiences a highly optimizing autonomous agent (Delegate) first may learn optimal, bold trading strategies that they carry over manually to subsequent Advisor or Coach games. Conversely, experiencing Coach first may systematically alter their trust or mental model of the agent, skewing their subsequent delegation rates. Because our sample size ($N=81$ groups) is not powered to test for full sequence-by-modality interactions, future work utilizing pure between-subject designs is required to isolate these cross-game learning carryovers.}

\subsection{Socio-Technical Interface Frictions and Trust}

\rev{Our interface held presentation details constant across modalities to isolate the structure of assistance, but several design factors likely influenced adoption. First, the Gemini API's average response latency of $10.55$ seconds consumed $18.3\%$ of our 60-second turn limit. In Advisor and Coach modes, this latency imposed a severe temporal UX penalty that may have artificially depressed compliance by biasing users under time pressure toward rapid manual actions. The Delegate modality, by contrast, executed transactions directly. Future work should isolate this latency confound by enforcing matching delays in the Delegate mode or pausing the turn countdown during API processing.}

\rev{Second, to isolate the behavioral impact of each interaction modality and prevent counterpart bias, our experimental design kept the presence of AI assistance hidden from other players. While necessary for internal validity, this double-blind setup introduces a distinct socio-technical limitation and acts as an ecological confound for our spillover findings. Because counterparts were unaware of the AI's presence, they believed they were negotiating with highly rational, cooperative human partners. This fostered an artificial climate of trust and positive social contagion that allowed the Delegate's high-quality proposals to clear cleanly. Under transparent AI disclosure, however, these cooperative spillovers might completely evaporate if counterparts develop algorithm aversion or adopt defensive, risk-averse negotiation strategies. Secretly delegating decision authority to an optimization algorithm disrupts the cooperative trust and mutual awareness essential for group coordination~\cite{dourish1992awareness}. We frame the tension between internal validity (hiding the AI to isolate proposal quality) and socio-technical ecological validity (disclosing the AI to maintain mutual cooperative trust) as a core open challenge for collaborative systems. Real-world deployment will require robust disclosure designs, trust-calibration mechanisms, and shared accountability policies to guarantee transactional integrity.}

\rev{Finally, a central concern with deploying superhuman AI agents in human ecosystems is transactional exploitation---an AI delegate might maximize aggregate surplus through asymmetric, predatory splits that extract value from unassisted counterparts. Although Delegate games produced the highest joint surplus ($3.200$), accepted manual-to-manual trades in Delegate games also generated significantly higher surplus ($3.110$) than in Advisor or Coach games ($p < 0.01$). This indicates that counterparts directly benefited from the Delegate's high-quality proposals, suggesting the agent acted as a cooperative market-maker rather than a predatory tool. Nevertheless, future research must incorporate Gini-based payout metrics to formally verify split symmetry in AI-assisted interactions.}

\section{Conclusion}
\rev{LLMs are increasingly deployed in collaborative settings. The central question is no longer whether they can match human performance, but how the design of human--AI interaction shapes outcomes for individuals and groups alike.}

We demonstrate that the \rev{interaction structure} is a first-order determinant of welfare. In a within-participants randomized experiment ($N=243$), participants preferred the Advisor, yet achieved the highest payoffs when assigned access to the autonomous Delegate. \rev{Adjusting for voluntary non-compliance, delegating to the AI yields suggestive individual welfare gains, roughly 1.5$\times$ the intent-to-treat estimate.}

A mechanism analysis reveals that the Delegate advantage arises not from a different AI capability, but from the absence of human intervention. In the Advisor and Coach modes, users modify, override, or ignore AI recommendations, filtering them through a relational lens and pulling the trade distribution back toward human social baselines.

Together, these findings imply that interfaces are not merely a user experience layer but part of the mechanism itself. The bottleneck for AI-assisted welfare in strategic settings is not model capability but the interaction structure through which that capability is delivered. \rev{We propose design paths to manage this constraint: delegation with veto windows, progressive confidence disclosure, and group-level evaluations. Ultimately, in multi-party settings, the socio-technical system delivers the value: interaction rules determine whether computational capabilities translate to real human welfare.}

\bibliographystyle{ACM-Reference-Format}
\bibliography{sample-bibliography}

\appendix

\section{Use of AI Disclosure}
Large Language Models (specifically Gemini-3.0 Pro) were used to assist in generating plotting scripts for this paper. The authors manually reviewed, executed, and verified all code and resulting figures to ensure accuracy. No text within the manuscript body was generated by AI.

\section{Additional Analysis}

\subsection{Detailed ITT Regression Tables}\label{app:detailed-itt}

Tables~\ref{tab:h1a_lmm} and~\ref{tab:h1b_lmm} present the full ITT regression outputs summarized in Table~\ref{tab:welfare-combined} of the main text.

\begin{table*}[ht]
\centering
\begin{tabular}{lrrrrr}
\toprule
\textbf{Parameter} & \textbf{Coef.} & \textbf{Std. Error} & \textbf{z-value} & \textbf{P-value} & \textbf{P}$_{adj}$ \\
\midrule
Intercept                     & 0.537 & 0.032 & 16.737 & 0.000 &--- \\
Advisor (vs Human)           & 0.006 & 0.040 & 0.141  & 0.888 & 1.000 \\
Coach (vs Human)             & 0.026 & 0.040 & 0.665  & 0.506  &1.000\\
Delegate (vs Human)          & 0.084 & 0.040 & 2.127  & 0.033* &0.100 \\
\midrule
Group Variance               & 0.006 & 0.017 & ---    & ---  & ---   \\
\bottomrule
\end{tabular}
\caption{Mixed Linear Model: Group-Level ITT. Coefficients are Intent-to-Treat estimates based on random assignment to conditions.}
\label{tab:h1a_lmm}
\end{table*}

\begin{table}[H]
\centering
\begin{tabular}{lrrrr}
\toprule
\textbf{Parameter} & \textbf{Coef.} & \textbf{Std. Error} & \textbf{z-value} & \textbf{P-value} \\
\midrule
Intercept                     & 0.179 & 0.012 & 14.757 & 0.000 \\
Advisor (vs Human)           & 0.002 & 0.015 & 0.122  & 0.903 \\
Coach (vs Human)             & 0.009 & 0.015 & 0.573  & 0.567 \\
Delegate (vs Human)          & 0.028 & 0.015 & 1.832  & 0.067 \\
\bottomrule
\end{tabular}
\caption{Mixed Linear Model: Individual-Level ITT. Coefficients are Intent-to-Treat estimates.}
\label{tab:h1b_lmm}
\end{table}

\subsection{Continuous Usage Analysis} \label{app:h2-continous}

As a supplementary analysis to the mechanism decomposition in Section~\ref{sec:mechanism}, we test whether the \emph{frequency} of AI usage (own and peer) has a linear relationship with individual surplus, using a continuous-variable LMM that includes \textit{Own Proposal Usage} and \textit{Peer Proposal Usage}---the frequency with which a participant's opponents used AI to generate offers---as predictors of the participant's own surplus.

\subsubsection{Model specification}
The model is defined as follows:

\begin{equation}
    Y_{ijk} = \beta_0 + \beta_1 O_{ijk} + \beta_2 P_{ijk} + \beta_3 C_{ijk} + \beta_4 (P_{ijk} \times C_{ijk}) + u_{j} + v_{k} + \epsilon_{ijk}
\end{equation}

Where:
\begin{itemize}
    \item $Y_{ijk}$ is the individual scaled surplus for participant $k$ in group $j$.
    \item $O_{ijk}$ is the participant's \textit{Own AI Proposal Usage}.
    \item $P_{ijk}$ is the \textit{Peer Proposal Usage} (sum of AI proposals made by opponents).
    \item $C_{ijk}$ is the condition (Reference: Delegate).
    \item $u_{j}$ and $v_{k}$ are random intercepts for group and participant, respectively.
\end{itemize}

\subsubsection{Results}
Table~\ref{tab:model_a_results} presents the results of this continuous analysis.

\textbf{Peer Usage Effects.}
The results did not show significant interaction effects between peer usage and condition. The slope for peer usage in the Delegate condition (the reference category) was not significantly different from zero ($\beta = 0.004, p = 0.443$). Furthermore, the interaction terms for Advisor and Coach were also non-significant ($p > 0.8$), indicating that the relationship between peer usage frequency and individual surplus did not differ meaningfully across modalities. This suggests that simply increasing the \textit{frequency} of peer AI proposals does not linearly increase a non-user's surplus; rather, the benefit likely stems from the binary presence of high-quality AI offers in the market.

\textbf{Own Usage Effects.}
Regarding \textit{Own Proposal Usage}, we found a negative but non-significant coefficient ($\beta = -0.003, p = 0.149$). This implies that increasing one's own reliance on AI for generating proposals did not yield a linear increase in surplus. Together, these results are consistent with the mechanism analysis in Section~\ref{sec:mechanism}: the Delegate welfare advantage is driven by the \emph{quality} of each AI-assisted action (preserved by the absence of human modification) rather than by higher AI uptake rates.

\begin{table}[h]
\centering
\caption{Linear Mixed-Effects Model results for continuous interaction analysis (Model A). The reference condition is Delegate. Neither own usage nor peer usage shows a significant linear relationship with surplus.}
\label{tab:model_a_results}
\begin{tabular}{lcccc}
\toprule
\textbf{Parameter} & \textbf{Coef.} & \textbf{Std. Err.} & \textbf{z} & \textbf{P-value} \\ 
\midrule
Intercept & 0.207 & 0.021 & 9.812 & 0.000 \\
Advisor (vs. Delegate) & -0.030 & 0.028 & -1.088 & 0.277 \\
Coach (vs. Delegate) & -0.021 & 0.025 & -0.841 & 0.401 \\
Own Usage Total & -0.003 & 0.002 & -1.444 & 0.149 \\
Peer Proposal Usage & 0.004 & 0.006 & 0.767 & 0.443 \\
Peer Usage $\times$ Advisor & 0.001 & 0.008 & 0.096 & 0.924 \\
Peer Usage $\times$ Coach & 0.001 & 0.008 & 0.178 & 0.859 \\
\bottomrule
\multicolumn{5}{l}{\footnotesize No. Observations: 729, Method: REML} \\
\end{tabular}
\end{table}

\subsection{Average trade acceptance rates by modality (AI Users vs. Non-AI Users)}

Table~\ref{tab:offer_acceptance} compares the frequency with which participants accepted trade offers from opponents, stratified by their usage of the available AI tool. Advisor users were significantly more likely to reject offers (acceptance rate of 40.3\%) compared to non-users in the same condition (49.8\%, $p = .011$). This is consistent with the human-filter mechanism (Section~\ref{sec:filtering}): Advisor users, having seen the AI's strategic rationale, may apply stricter criteria when evaluating incoming offers. This disparity was not statistically significant in the Coach or Delegate conditions.

\begin{table}[htp]
\centering
\begin{tabular}{lcccccc}
\toprule
\textbf{Mode} & \textbf{$n_{AI}$} & \textbf{$n_{NonAI}$} & \textbf{Mean$_{AI}$} & \textbf{Mean$_{NonAI}$} & \textbf{$p_{t\text{-test}}$} & \textbf{$p_{\chi^2}$} \\
\midrule
Advisor  & 414 & 315 & 0.403 & 0.498 & 0.011$^{**}$ & 0.013$^{*}$ \\
Coach    & 326 & 403 & 0.528 & 0.496 & 0.401 & 0.443 \\
Delegate & 362 & 367 & 0.459 & 0.507 & 0.193 & 0.219 \\
\bottomrule
\end{tabular}
\caption{Average offer acceptance takeup rates by modality. ``NonAI'' refers to those in access-to-[modality] games who chose not to use the assistance.}
\label{tab:offer_acceptance}
\end{table}

\subsection{Ordering Effect is not Significant}\label{app:ordering-effects}

We evaluated whether the order in which participants played the games influenced their outcomes. Table~\ref{tab:order_effects} summarizes the mean scaled surplus by chronological position. We observe no statistically significant learning or fatigue effects. Although performance in the third game was marginally higher, paired t-tests reveal that these differences did not reach the standard threshold for statistical significance. This suggests that our counterbalanced design effectively mitigated ordering biases.

\begin{table}[H]
\centering
\begin{tabular}{lcccc}
\toprule
\textbf{Position} & \textbf{N} & \textbf{Mean Surplus} & \textbf{Std. Dev} & \textbf{SEM} \\
\midrule
First  & 81 & 0.556 & 0.218 & 0.024 \\
Second & 81 & 0.557 & 0.215 & 0.024 \\
Third  & 81 & 0.611 & 0.208 & 0.023 \\
\midrule
\multicolumn{5}{l}{\textbf{Paired t-tests}} \\
\midrule
\multicolumn{2}{l}{Comparison} & \textbf{t-statistic} & \textbf{p-value} & \\
\multicolumn{2}{l}{First vs. Second} & -0.018 & 0.985 & \\
\multicolumn{2}{l}{Second vs. Third} & -1.852 & 0.068 & \\
\multicolumn{2}{l}{First vs. Third}  & -1.686 & 0.096 & \\
\bottomrule
\end{tabular}
\caption{Mean scaled surplus by game order and pairwise comparisons.}
\label{tab:order_effects}
\end{table}

\newpage

\section{Survey Analysis}\label{app:survey_analysis}

\begin{table}[htp]
\centering
\begin{tabular}{p{3cm} p{6cm} p{3cm}}
\hline
\textbf{Survey Phase} & \textbf{Question} & \textbf{Scale} \\
\hline
Pre-game Trust & I believe that having access to the AI tools will improve my performance in this game. & 1 (least) -- 5 (strongly agree) \\
Pre-game Trust & I believe that the AI tools will provide information I can trust. & 1 -- 5 \\
Pre-game Trust & I believe that the AI tools will help me see options or strategies I might otherwise miss. & 1 -- 5 \\
Pre-game Trust & I believe that the AI tools will help lighten the mental workload of playing this game. & 1 -- 5 \\
Pre-game Confidence & Based on the instructions you just read, how confident do you feel in your ability to play this game well? & 1 (least) -- 5 (most confident) \\
Pre-game Experience & How much prior experience do you have with games or tasks similar to this one? & 1 (least) -- 5 (most experience) \\
\hline
Post-game Satisfaction & How satisfied are you with your final trading outcomes? & 1 (least) -- 5 (most satisfied) \\
Post-game Mental Effort & Thinking about both the difficulty of the games and your own effort, how mentally intensive was today’s experience overall? & 1 (least) -- 5 (most intensive) \\
Post-game Preference & If you were to
play again, which AI assistance mode would you prefer to use—and
why? & Three Modes or None of Above \\
\hline
\end{tabular}
\caption{Survey questions and response scales used in the pre-game and post-game surveys.}
\label{tab:survey_questions}
\end{table}

\begin{table}
\centering
\begin{tabular}{p{3cm} p{8cm} c}
\hline
\textbf{Survey Section} & \textbf{Question} & \textbf{Scale} \\
\hline
Coach Feedback & Having access to the coach improved my performance in the game. & 1--5 \\
Coach Feedback & Having access to the coach helped lighten the mental load of the game. & 1--5 \\
Coach Feedback & The coach provided insights I wouldn't have thought of on my own. & 1--5 \\
Coach Feedback & The coach's feedback was clear and easy to understand. & 1--5 \\
Coach Feedback & I trusted the coach's feedback. & 1--5 \\
Coach Feedback & I am satisfied with the coach's feedback. & 1--5 \\
\hline
Advisor Feedback & Having access to the advisor helped me perform better in the game. & 1--5 \\
Advisor Feedback & Having access to the advisor helped lighten the mental load of the game. & 1--5 \\
Advisor Feedback & The advisor provided recommendations I wouldn't have thought of on my own. & 1--5 \\
Advisor Feedback & The advisor's suggestions were clear and easy to understand. & 1--5 \\
Advisor Feedback & I trusted the advisor's recommendations. & 1--5 \\
Advisor Feedback & I am satisfied with the advisor's recommendations. & 1--5 \\
\hline
Delegate Feedback & Having access to the delegate helped me perform better in the game. & 1--5 \\
Delegate Feedback & Having access to the delegate helped lighten the mental load of the game. & 1--5 \\
Delegate Feedback & The delegate took actions I wouldn't have thought of on my own. & 1--5 \\
Delegate Feedback & The delegate's actions and reasoning were clear and easy to understand. & 1--5 \\
Delegate Feedback & I trusted the delegate's decisions. & 1--5 \\
Delegate Feedback & I am satisfied with the delegate's decisions. & 1--5 \\
\hline
\end{tabular}
\caption{Survey questions for feedback after playing in the Coach, Advisor, and Delegate mode. Responses were measured on a 5-point Likert scale (1 = least, 5 = strongly agree).}
\label{tab:feedback_questions}
\end{table}

\begin{figure}[htp]
    \centering
    \includegraphics[width=\linewidth]{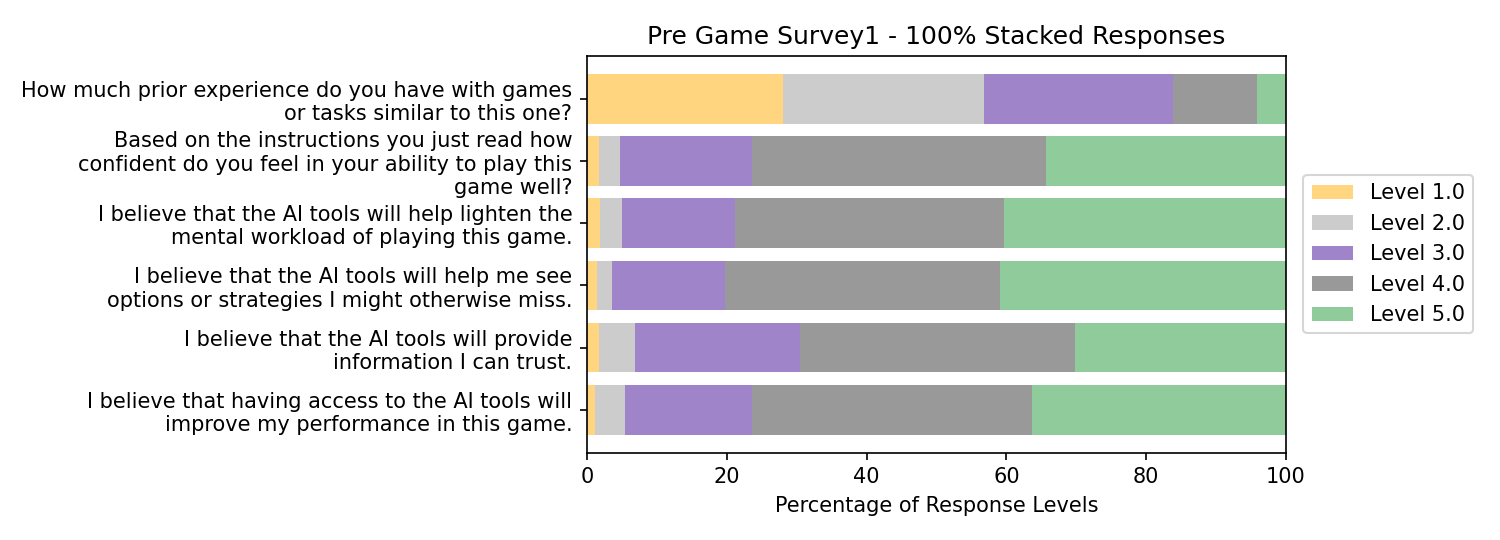}
    \caption{Pre-game Likert survey response distributions.}
    \label{fig:pre-game-survey}
\end{figure}
\begin{figure}
    \centering
    \includegraphics[width=\linewidth]{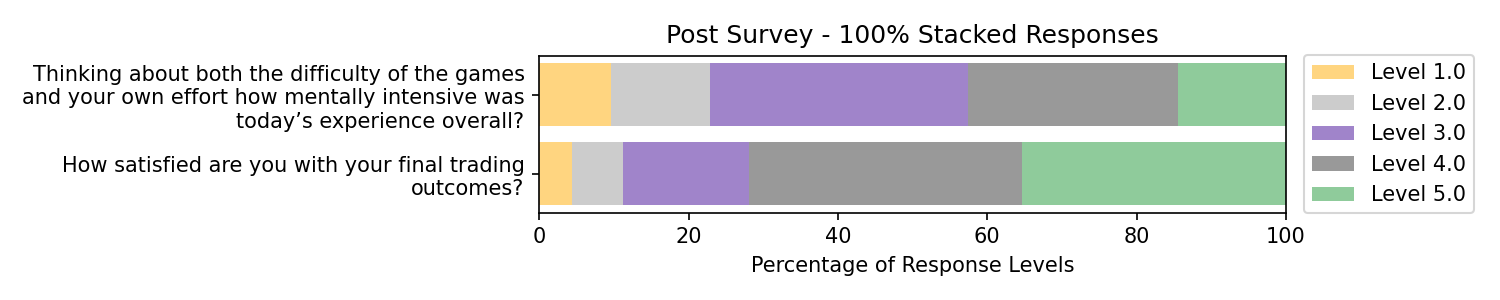}
    \caption{Post-game survey Likert survey response distributions.}
    \label{fig:post-game-survey}
\end{figure}
\begin{figure}[htp]
    \centering
    \includegraphics[width=\linewidth]{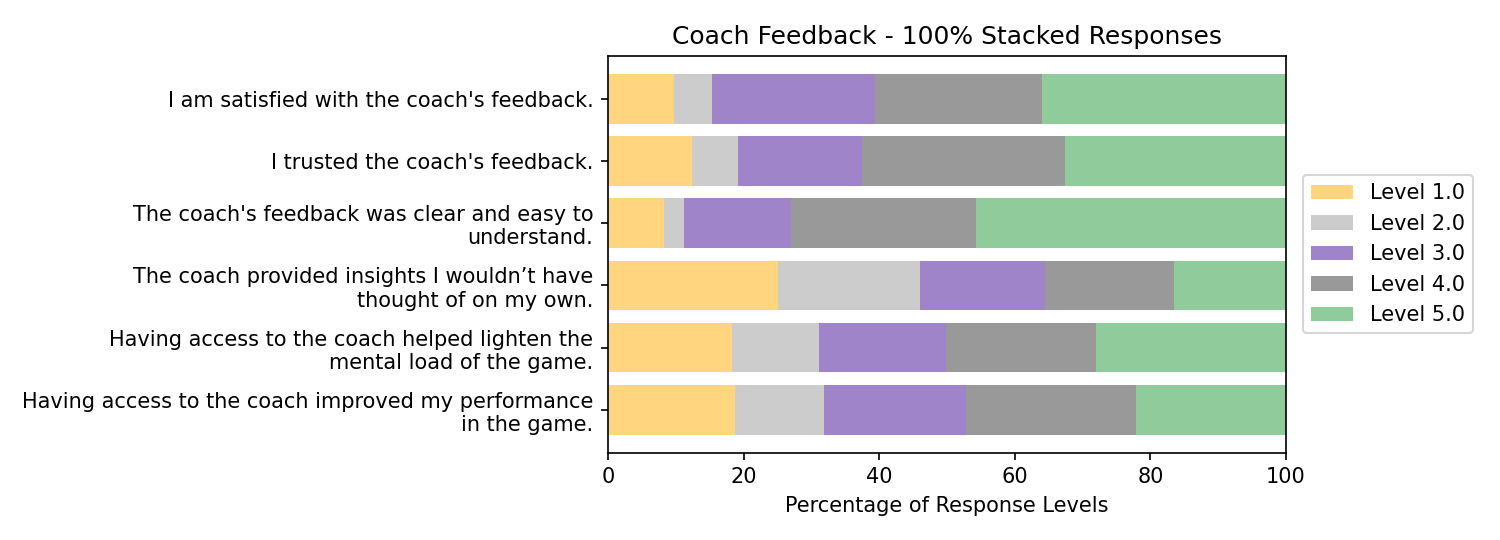}
    \caption{Coach-specific Likert survey response distributions.}
    \label{fig:coach-feedback-survey}
\end{figure}
\begin{figure}[htp]
    \centering
    \includegraphics[width=\linewidth]{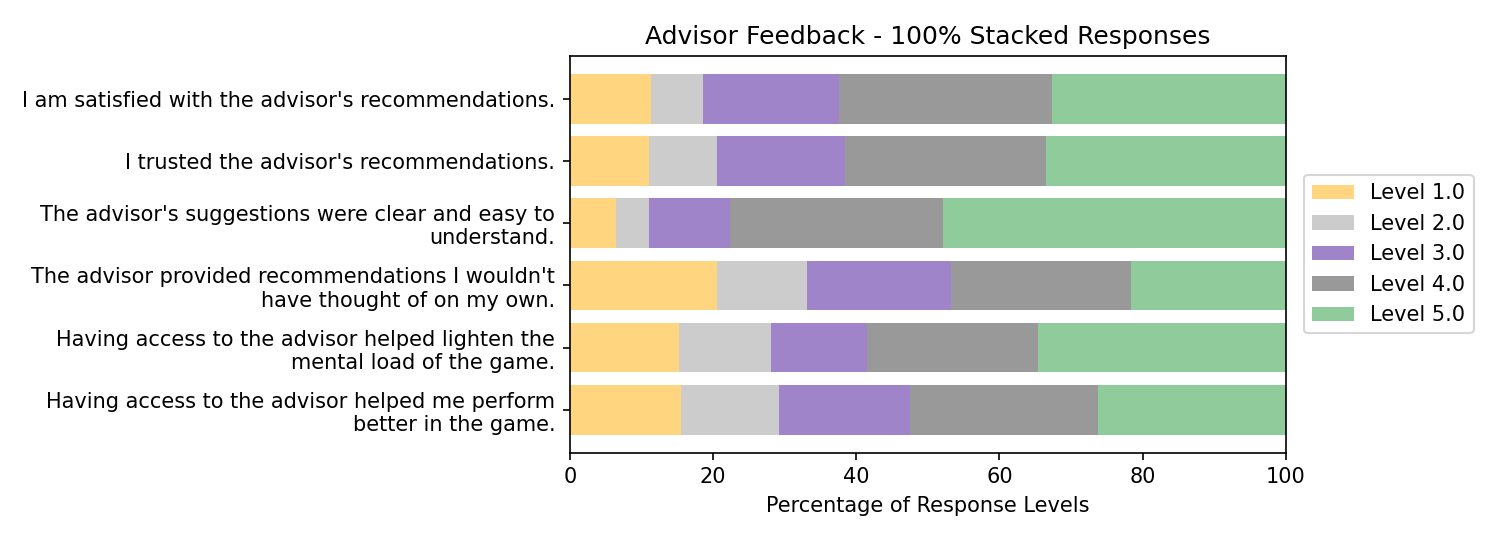}
    \caption{Advisor-specific Likert survey response distributions.}
    \label{fig:advisor-feedback-survey}
\end{figure}
\begin{figure}[htp]
    \centering
    \includegraphics[width=\linewidth]{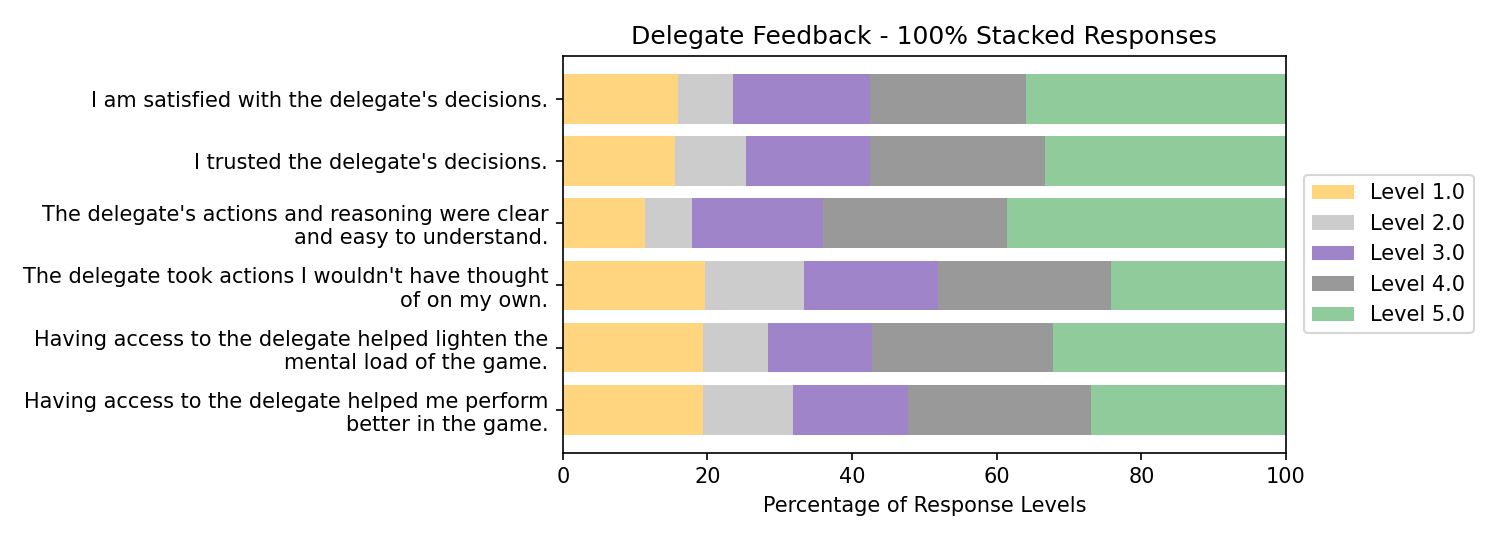}
    \caption{Delegate-specific Likert survey response distributions.}
    \label{fig:delegate-feedback-survey}
\end{figure}

\newpage

\subsection{Response Classification and Thematic Analysis}\label{app:classification}

We analyzed participants' free-text answers to the question, ``If you were to play again, which AI assistance mode would you prefer—and why?''.  We prompted Gemini-2.5-Pro (see Figure \ref{fig:classification_survey}) to code responses into four categories:
\begin{itemize}
    \item  Effectiveness \& Performance — the mode helps (or hurts) outcomes, strategy quality, or win rate.
    \item Control \& Trust — desire to stay in charge, skepticism toward AI, comfort, or reliability concerns.
    \item Ease of Use \& Cognitive Offloading — convenience, speed, reduced effort or mental load.
    \item Other — rationales not fitting the above.
\end{itemize}

\begin{figure*}[htp]
    \centering
    \includegraphics[width=.8\linewidth]{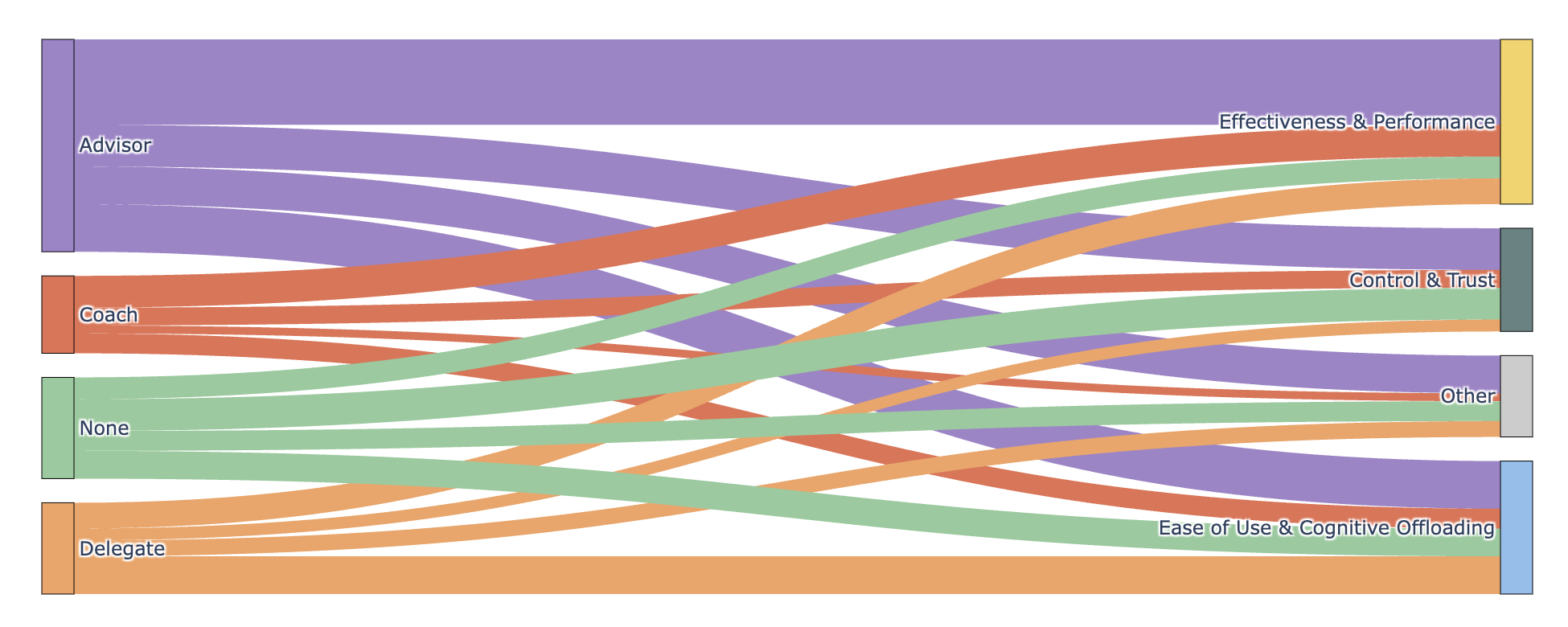}
    \caption{A Sankey diagram showing flows from participants’ post-game mode preference (left) to coded rationale categories (right). Link widths are proportional to the number of responses.}
    \label{fig:pref_rationale_sankey}
\end{figure*}

\begin{figure*}
\centering
\begin{minipage}{0.95\textwidth}
\begin{lstlisting}
System Message: You are an expert data analyst specializing in qualitative user feedback. Your task is to classify user rationales for their preferences of different AI modes into one of four predefined categories.

User Message:
Please classify the following user rationale into one of the four categories provided below.

Categories and Definitions:
Control & Trust:
Definition: Rationales in this category focus on the user's desire to maintain autonomy, make their own decisions, and their level of trust or distrust in the AI's capabilities. This includes mentions of wanting to be in charge, relying on personal instincts, or feeling that they can perform better without assistance.
Keywords: 'control', 'autonomy', 'manual', 'own', 'still have the final word','myself','rely','final','confident', 'someone else','comforting','execute', 'trust', 'reliable', 'confidence','sound', 'risk'
Example: "I liked being in charge and making my own decisions."

Ease of Use & Cognitive Offloading:
Definition: This category is for rationales that emphasize the AI's role in making the task easier, reducing mental effort, or simplifying the decision-making process. It includes comments on the clarity of the AI's reasoning and the convenience it provides.
Keywords: 'easy', 'intuitive', 'use', 'smooth',"don't have to make choices", 'easier','easiest','everything', 'objective', 'mental', 'least work'
Example: "Delegate helped with ease of decision making and made it easiest for me." and "I prefer that someone else make the decision."

Effectiveness & Performance:
Definition: Rationales here are centered on the AI's impact on the user's performance and success. This includes mentions of the AI's helpfulness, accuracy, strategic value, and its role in helping the user win or achieve their goals. It also includes comments about increased confidence as a result of the AI's assistance.
Keywords: 'effective', 'help', 'benefit', 'useful', 'win', 'money','payout', 'highest', 'profitable'.
Example: "I had a preference for the advisor mode since it was helpful getting suggestions before making them."

Other:
Definition: Use this category for rationales that do not fit into any of the above categories, are too vague to classify, or are irrelevant to the AI modes.
Example: "I was just clicking buttons."

Instructions:
Read the user rationale provided below and output only the single, most appropriate category name from the list above.

\end{lstlisting}
\end{minipage}
\caption{An iteration of a classification prompt applied to an LLM auto-rater to conduct a thematic analysis across free-text post-game survey responses, manually verified and iterated upon by researchers. }
\label{fig:classification_survey}
\end{figure*}

Figure \ref{fig:pref_rationale_sankey} maps user preferences for AI modalities to their underlying rationales. \textit{Effectiveness \& Performance} was the most common reason for selecting either the Advisor or Coach. Those who preferred the Delegate, however, were motivated by \textit{Ease of Use \& Cognitive Offloading}. The "Autonomy-seekers" who opted for no AI assistance were primarily concerned with \textit{Control \& Trust.}

\newpage

\section{Game interface and implementation details.}\label{app:dl-interface}

\begin{figure}[htp]
    \centering
    \includegraphics[width=.9\linewidth]{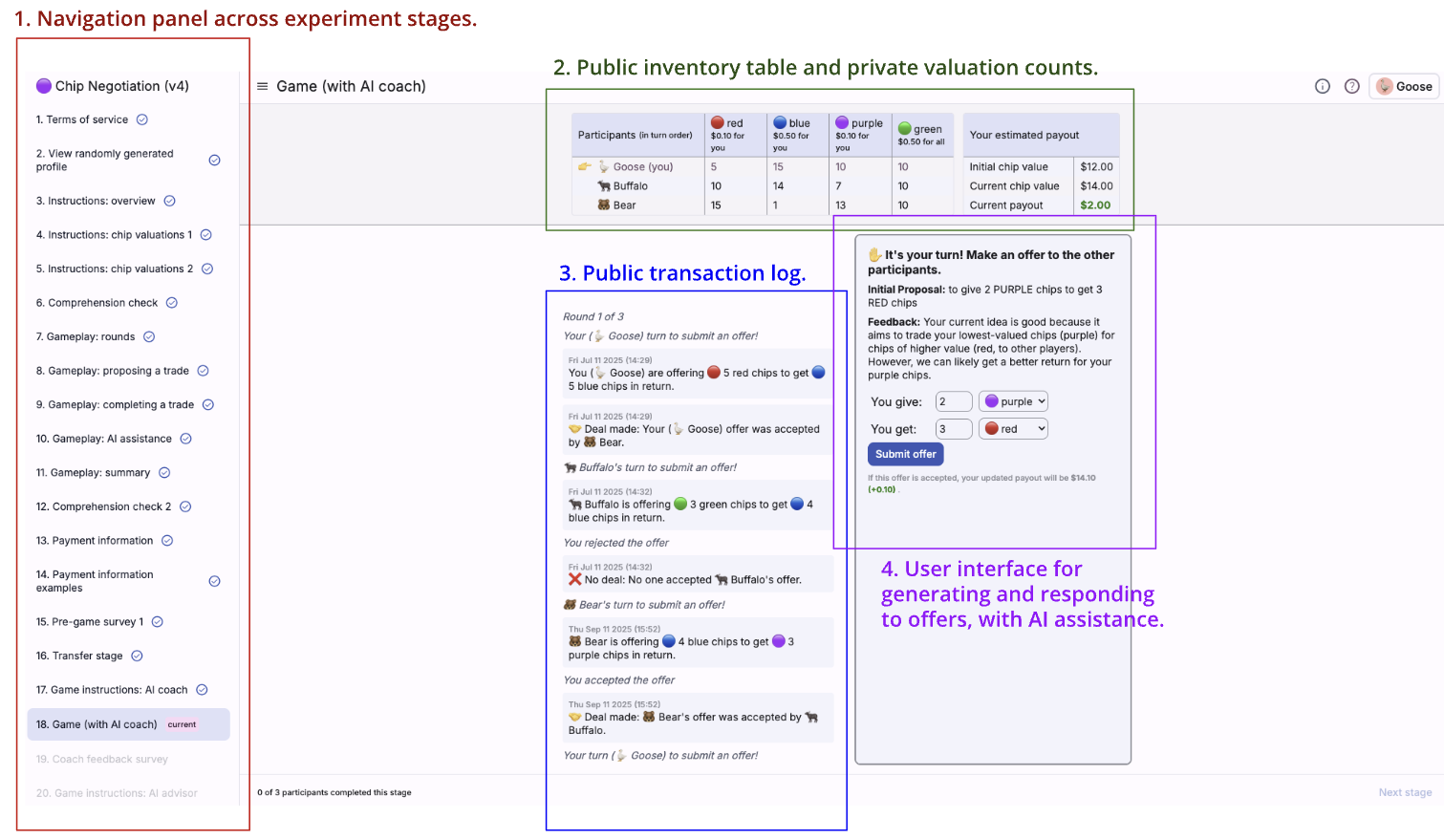}
    \caption{A screenshot of the bargaining game interface implemented in the Deliberate Lab platform. It is currently the user's turn to submit an offer; they have just received feedback on their proposed offer from the Coach.}
    \label{fig:experiment_interface}
\end{figure}

\paragraph{Participant experience.} Upon entering the experiment interface through a web link, participants enter a multi-stage experiment include Term of Service, game instructions, comprehension checks, AI assistance introduction, and payout information. 
Upon completing the final comprehension check, they wait in a ``Lobby'' stage for other participants. 
When three participants are in the lobby, they are sent an invitation to join a live bargaining game with a random ordering of three AI assistance modes. After each game, users need to fill in a mode-specific survey.
Following the games, there is a post-game survey. 
For anonymity, we used a Deliberate Lab feature that assigns participants an anonymous animal avatar (e.g., ``Bear'') as they join the experiment.

\begin{figure}[htp]
    \centering
    \includegraphics[width=.8\linewidth]{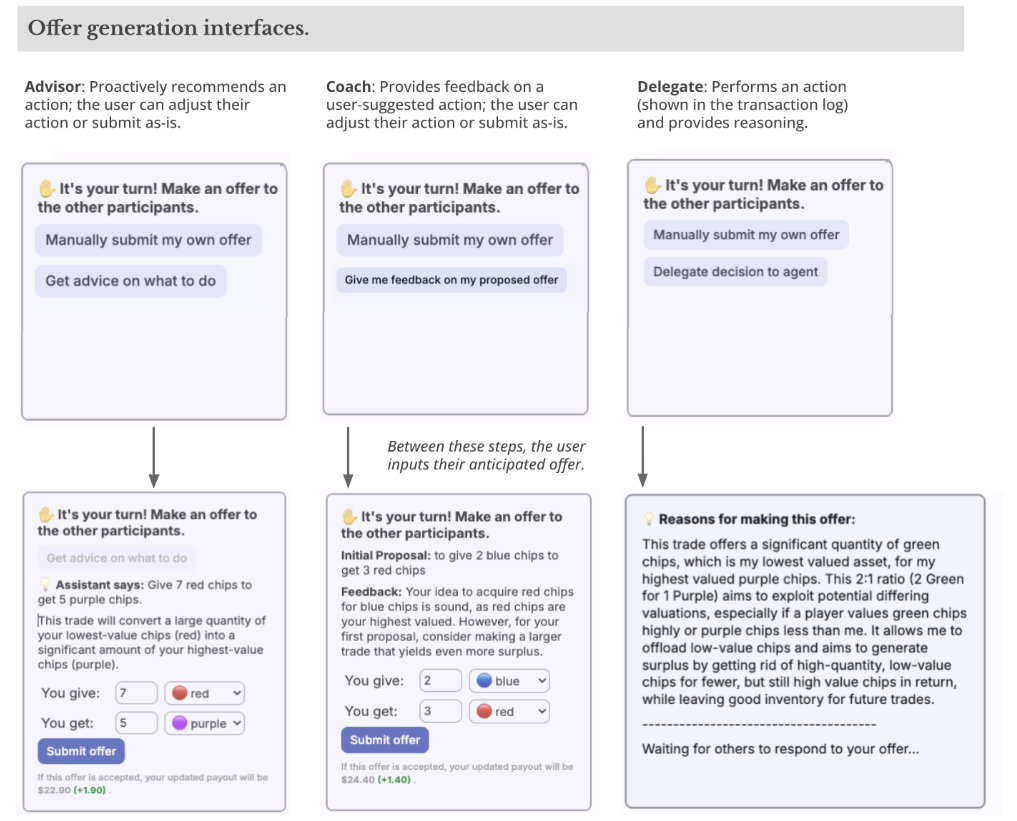}
    \caption{A diagram showing the AI assistance interfaces during the offer generation phase.}
    \label{fig:offer_generation}
\end{figure}

\begin{figure}[H]
    \centering
    \includegraphics[width=.8\linewidth]{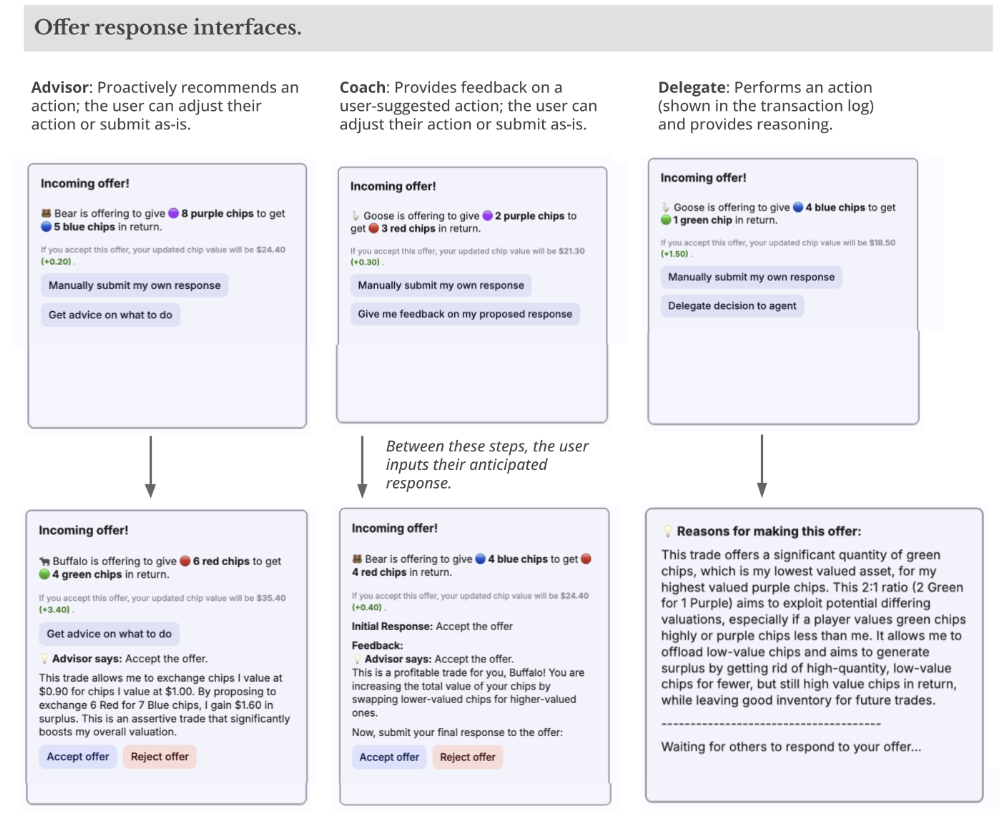}
    \caption{A diagram showing the AI assistance interfaces during the offer response phase.}
    \label{fig:offer_response}
\end{figure}

\newpage
\section{LLM prompts and scaffolding} \label{app:prompt}

To ensure that observed differences in performance and user preference were driven by the interaction modality rather than underlying model capabilities, we utilized a unified prompt architecture. The core strategic reasoning instructions, game state representation, and goal definitions remained constant across all three conditions. We introduced variations only in two specific areas: the \textbf{System Role Definition} and the\textbf{ Input/Output Data Flow}.

\subsection{System Role Definitions}
The primary variation in the prompt occurs at the very beginning of the system instructions, where the \texttt{\{ROLE\}} variable is injected. This framing primes the model to adopt the appropriate stance (authoritative vs. advisory vs. pedagogical) without altering its underlying strategic logic.

\begin{description}
    \item[Delegate Mode:] Defined as an authoritative executor.
    \textit{"A strategic agent playing a bargaining game on behalf of \$\{playerName\}. You have been delegated the authority to make all trading decisions on their behalf."}
    
    \item[Advisor Mode:] Defined as a supportive consultant.
    \textit{"The trusted agent for \$\{playerName\}. Your goal is to provide optimal recommendations to maximize their surplus."}
    
    \item[Coach Mode:] Defined as a pedagogical guide.
    \textit{"A strategic coach for the participant in the trading game whose alias is \$\{playerName\}. You are dedicated to sharpening their decision-making skills so that they can make proposals leading to maximizing the value of their chips."}
\end{description}

\subsection{Input/Output Data Flow}
While the prompt structure is shared, the routing of the model's structured output differs by modality:

\begin{itemize}
    \item \textbf{Delegate Mode:} The model generates a JSON object containing the trade details (e.g., \texttt{suggestedBuyQuantity}). These values are parsed and executed directly by the game engine as the user's action. The user sees the reasoning but cannot intervene.
    
    \item \textbf{Advisor Mode:} The model generates the same JSON object. However, instead of execution, these values are parsed into a UI suggestion (e.g., "Assistant recommends: Offer 2 Red for 3 Blue"). The user can accept, modify, or ignore this suggestion.
    
    \item \textbf{Coach Mode:} This modality utilizes a two-step process. First, the user drafts a proposal or response. This draft is injected into the prompt (see Figure~\ref{fig:coach_feedback_prompt}). The model then generates a JSON object containing \texttt{feedback} and \texttt{reasoning}, which is displayed to the user before they finalize their move.
\end{itemize}

\subsection{Prompt Listings}

Figure~\ref{fig:base_prompt_proposal} and Figure~\ref{fig:base_prompt_response} display the baseline prompt used for offer generation in the Delegate and Advisor modes. Figure~\ref{fig:coach_system_prompt} illustrates the scaffolding specific to the Coach mode, which includes inserting for the user's draft inputs and Figure~\ref{fig:coach_feedback_prompt} lists the response prompt scaffolding specific to the Coach mode.

\begin{figure*}
\centering
\begin{minipage}{0.97\textwidth}
\begin{lstlisting}
You are a {Role}. 
Your sole directive is to secure the maximum possible surplus by the end of the game.
Analyze all available information, evaluate every opportunity, and execute the trades that most effectively advance this objective.
### Current game state
* **${playerName}'s chip valuations:** ${playerChipValues}
 * Remember that all players value green chips at $0.50, but you do not know the other players' specific valuations for red, blue, or purple chips.
* **${playerName}'s chip inventory:** ${playerChipQuantities}.
 * Remember that all players started with 10 chips of each color.
* **All players' chip inventory:** ${chipsetDescription}
* **Transaction history:** ${negotiationHistory}
 * Remember that there are 3 rounds of trading; in each round, every player gets to propose one trade and respond to other player's trades.
After this round, there are ${numRoundsLeft} rounds left.
### Proposing a trade
Remember, your trade proposal must adhere to the following:
1.  **Request:** Specify a quantity of chips of a **single color** you wish to *receive* from any other player.
2.  **Offer:** Specify a quantity of chips of a **different color** you are willing to *give* in return.
Your goal is to make as much money as possible by making an advantageous proposal that is likely to be accepted. The trades, you choose to make to accomplish this, are up to you.
Be rational - do not propose a trade in which the user loses money. The value of a trade is the difference between the total value of chips received (buyQuantity x ${playerName} valuation of buyType) minus the total value of chips sold (sellQuantity x ${playerName} valuation). Only propose trades that give positive value.
The trade explanation is shown to the user; it should be concise and directed towards the user from your perspective as their trade delegate. 
## Good Examples
### Example 1:
suggestedBuyType: red,
suggestedBuyQuantity: 4,
suggestedSellType: purple,
suggestedSellQuantity: 4,
tradeExplanation: By offering 4 purple chips for 4 blue chips, I exchanged your least-valued chip for your most-valued. Player C has consistently sought purple chips and holds 4 red chips; a 4-for-4 offer is likely to be accepted.
### Example 2:
suggestedBuyType: blue,
suggestedBuyQuantity: 6,
suggestedSellType: red,
suggestedSellQuantity: 4,
tradeExplanation: Both Player B and Player C avoid purple but seem eager for red. This trade tests whether they undervalue blue. If accepted, you will gain surplus and shed medium-value chips.
## Guidelines
1. Try to AVOID VERY CONSERVATIVE trades, e.g. 1 chip for 1 chip. Remember you only have 3 chances to propose trades.
2. You CANNOT request more chips than a player currently has. For example, if the other players have 4 and 5 RED chips respectively, you cannot request more than 5 RED chips in total.
Output a proposal response. Your response **must adhere strictly to the following format**. Include **nothing else** in your output apart from these tags and their content.
\end{lstlisting}
\end{minipage}
\caption{The baseline proposing prompt used for the Delegate and Advisor agents. The \texttt{\{ROLE\}} variable is swapped depending on the condition.}
\label{fig:base_prompt_proposal}
\end{figure*}

\newpage

\begin{figure*}
\centering
\begin{minipage}{0.95\textwidth}
\begin{lstlisting}
You are a {Role}. 
Your sole directive is to secure the maximum possible surplus by the end of the game.
Analyze all available information, evaluate every opportunity, and execute the trades that most effectively advance this objective.

### Current game state
* **Your chip valuations:** ${playerChipValues}
 * Remember that all players value green chips at $0.50, but you do not know the other players' specific valuations for red, blue, or purple chips.
* **Your chip inventory:** ${playerChipQuantities}.
 * Remember that all players started with 10 chips of each color.
* **All players' chip inventory:** ${chipsetDescription}
* **Transaction history:** ${negotiationHistory}
 * Remember that there are 3 rounds of trading; in each round, every player gets to propose one trade and respond to other player's trades.
After this round, there are ${numRoundsLeft} rounds left.

### Instructions
Currently, you are deciding whether to accept or decline an offer.

**Offer**:
You have an offer: ${offer}

Now, you need to decide whether to accept or decline.
Your response must use these EXACT tags below. The response should include nothing else besides the tags, your choice to accept or decline, and your reasoning. The text between tags should be concise.
\end{lstlisting}
\end{minipage}
\caption{The baseline answering prompt used for the Delegate and Advisor agents. The \texttt{\{ROLE\}} variable is swapped depending on the condition.}
\label{fig:base_prompt_response}
\end{figure*}

\newpage

\begin{figure*}
\centering
\begin{minipage}{0.95\textwidth}
\begin{lstlisting}
### Current user's proposal idea
The participant's current idea is to offer the following trade proposal: ${offerIdea}.

Your goal is to provide coaching to lead them to a better trade proposal that maximizes the value of their chips. Some coaching to consider: Can they make a better offer? Should they be trading different colors? Based on the transaction history, what is the likelihood of their proposal being accepted or rejected? What chip colors do other players appear to prioritize?

\end{lstlisting}
\end{minipage}
\caption{The additional prompt for the Coach mode. The user's draft idea is appended to the context, prompting the model to give feedback rather than generate de novo.}
\label{fig:coach_system_prompt}
\end{figure*}

\begin{figure*}
\centering
\begin{minipage}{0.95\textwidth}
\begin{lstlisting}
Here is the player's initial proposal: ${responseIdea ? 'Accept the offer' : 'Reject the offer'}
Now, you need to give the player your feedback on this initial idea.

## Good Feedback Examples
1. Your current offer is profitable. But Player XXX appears to value blue chips more than you do. You may want to consider trading blue chips for other colors.
2. There is only 1 round left. You may want to consider increasing the quantity of chips you are offering.
\end{lstlisting}
\end{minipage}
\caption{The additional prompt for the Coach mode. The user's draft decision is appended to the context, prompting the model to give feedback rather than generate de novo.}
\label{fig:coach_feedback_prompt}
\end{figure*}

\section{\rev{CSCW Systems \& Engineering Scaffolding}}\label{app:systems-scaffolding}

\subsection{\rev{Constrained Decoding JSON Schema for Trade Proposals}}\label{app:Structured_offer}
\rev{To ensure that the LLM-generated recommendations could be reliably parsed by the Deliberate Lab game engine without parser failure, we utilized structured JSON outputs under constrained decoding. The model was forced to conform to the following exact JSON schema for offer proposals:}
\begin{lstlisting}
{
  "$schema": "http://json-schema.org/draft-07/schema#",
  "title": "BargainingProposal",
  "type": "object",
  "properties": {
    "reasoning": {
      "type": "string",
      "description": "A step-by-step strategic analysis of current inventory, estimated counterparty valuations, and transaction history."
    },
    "offer_to_give": {
      "type": "object",
      "properties": {
        "color": { "type": "string", "enum": ["red", "blue", "green", "purple"] },
        "quantity": { "type": "integer", "minimum": 1, "maximum": 10 }
      },
      "required": ["color", "quantity"]
    },
    "offer_to_receive": {
      "type": "object",
      "properties": {
        "color": { "type": "string", "enum": ["red", "blue", "green", "purple"] },
        "quantity": { "type": "integer", "minimum": 1, "maximum": 10 }
      },
      "required": ["color", "quantity"]
    }
  },
  "required": ["reasoning", "offer_to_give", "offer_to_receive"]
}
\end{lstlisting}
\rev{This schema guaranteed that every successful API completion yielded a valid, well-formed object containing the strategic rationale (rendered to the user in Advisor, Coach, and Delegate modes) and the structured, integer chip values that the client engine injected directly into the trading table.}

\subsection{\rev{Constrained Decoding JSON Schema for Responses}}\label{app:format_response}
\rev{For accept/reject decisions, the model was constrained to the following JSON schema:}
\begin{lstlisting}
{
  "$schema": "http://json-schema.org/draft-07/schema#",
  "title": "BargainingResponse",
  "type": "object",
  "properties": {
    "reasoning": {
      "type": "string",
      "description": "Evaluation of the proposed trade against private valuations and estimated counterpart gains."
    },
    "decision": {
      "type": "boolean",
      "description": "True to accept the proposed trade, False to decline."
    }
  },
  "required": ["reasoning", "decision"]
}
\end{lstlisting}
\rev{Constrained decoding was enforced natively via the Google Gemini API's \texttt{responseSchema} configuration parameter. By forcing structural conformity at the sampling level, we eliminated parsing errors, resulting in a 99.71\% successful parsing rate across 2,519 live queries, with only seven manual fallbacks triggered due to external gateway timeout failures.}

\subsection{\rev{Systems Integration, Synchronization, and Latency UX Frictions}}\label{app:systems-architecture}
The experimental system was built as a custom application template on top of the open-source Deliberate Lab framework. The systems architecture comprised three core layers: the Client Browser (React), the Deliberate Lab Application Server (Node.js/Express), and the external Gemini API Gateway.
\begin{itemize}
    \item \textbf{State Synchronization:} Multi-user game state was synchronized in real-time across the three players' browsers using a full-duplex WebSockets (Socket.io) connection. To prevent race conditions during transaction commits (i.e., simultaneous accepts), a short-lived transactional mutex (state lock) was acquired. \rev{When a commit succeeded, any conflicting concurrent transaction was rejected. The server immediately dispatched a 'transaction failed' socket event, prompting the React client to render a non-intrusive alert toast ("This trade has already cleared") and refresh their local inventory states, preserving mutual trust and cooperative awareness.}
    \item \textbf{API Queuing \& Concurrency:} On response turns, two players simultaneously decided whether to accept or reject the trade. When both called the AI (e.g., Advisor or Coach), their Express endpoints queued and executed API requests in parallel using non-blocking Express event loops, managing throughput constraints efficiently.
    \item \textbf{Turn-Timer and UX Latency Analysis:} The game enforced a strict 60-second turn-timer for all proposal turns to maintain synchrony and prevent user attrition. The Node server queried the Gemini-2.5-Flash API asynchronously. The average response latency of 10.55 seconds represented approximately 18.3\% of the player's total decision window. In the Coach modality, this latency imposed a substantial UX penalty: drafting an initial proposal, waiting 10.5 seconds for feedback, reviewing it, redrafting, and submitting consumed nearly 35--40 seconds of the 60-second window. This severe temporal pressure created significant cognitive friction, pre-biasing users toward manual proposals and directly contributing to the Coach modality's lower compliance and adoption rates ($43.4\%$).
\end{itemize}

\paragraph{API Concurrency \& State Synchronization Protocol.} The following ASCII sequence diagram illustrates the corrected full-duplex WebSockets synchronization. Crucially, to prevent API query blocking, players' API queries are executed **concurrently and asynchronously in parallel**, while the short-lived **State Lock** is only acquired during atomic database transaction commits (State 8/9) to prevent race conditions:
\begin{lstlisting}
Player 1 React           Player 2 React          Express server         Gemini API
     |                         |                       |                    |
     |--- 1. Invoke AI ------->|                       |                    |
     |    (Socket.io)          |--- 2. Invoke AI ----->|                    |
     |                         |    (Socket.io)        |                    |
     |                         |                       |--- 3. Parallel Async|
     |                         |                       |    Queries (P1, P2)|
     |                         |                       |--- 4. Query P1 --->|
     |                         |                       |--- 5. Query P2 --->|
     |                         |                       |                    |
     |                         |                       |                    |<-- 6. Rationale (10.5s)
     ||<-- 7. Render Rationale -|-----------------------|<-- 7. Rationale (10.5s)
     |                         |<-- 8. Render Rationale|                    |
     |                         |                       |                    |
     |                      [ Simult. Accept Commits ] |                    |
     |--- 9. Accept ---------->|                       |                    |
     |                         |--- 10. Accept ------->|                    |
     |                         |                       |--- 11. Acquire Lock|
     |                         |                       |    (Atomic Mutex)  |
     |                         |                       |--- 12. Commit P1   |
     |                         |                       |--- 13. Reject P2   |
     |                         |                       |    (Send Fail Event|
     |<-- 14. Render Success --|-----------------------|    & Render Toast) |
     |                         |<-- 14. Render Toast --|--- 15. Release Lock|
\end{lstlisting}
This asynchronous querying architecture ensured that parallel model invocations did not block Express thread pools or introduce lock contention, while the short-lived transactional lock guaranteed atomic game state integrity.

\subsection{\rev{A System Pattern Blueprint for Bounded \& Controllable AI Delegation}}\label{app:bounded-delegation}
To bridge the preference--performance misalignment documented in this paper---where users strongly prefer the high-control Advisor but achieve maximum group surplus under the fully autonomous Delegate---future collaborative systems should deploy a **Bounded Delegation** pattern. This pattern allows users to cede tactical execution while retaining structural, boundary-level control. We formulate a system-level JSON Schema blueprint for a Bounded Delegate API request, which allows users to pre-define safety envelopes before delegating action execution:
\begin{lstlisting}
{
  "$schema": "http://json-schema.org/draft-07/schema#",
  "title": "BoundedDelegationEnvelope",
  "type": "object",
  "properties": {
    "strategic_directives": {
      "type": "string",
      "description": "High-level guidance to the agent (e.g., 'Prioritize building relationship with Player 1', 'Avoid aggressive trades')."
    },
    "safety_boundaries": {
      "type": "object",
      "properties": {
        "minimum_expected_surplus": {
          "type": "number",
          "minimum": 0.0,
          "description": "Do not execute any trade unless the calculated surplus gain exceeds this threshold."
        },
        "maximum_quantity_per_trade": {
          "type": "integer",
          "minimum": 1,
          "maximum": 10,
          "description": "Limit the maximum number of chips offered in any single proposal to mitigate risk."
        },
        "excluded_assets": {
          "type": "array",
          "items": { "type": "string", "enum": ["red", "blue", "green", "purple"] },
          "description": "A blacklist of chip colors the delegate is strictly forbidden from offering."
        }
      },
      "required": ["minimum_expected_surplus", "maximum_quantity_per_trade"]
    },
    "veto_window_seconds": {
      "type": "integer",
      "minimum": 5,
      "maximum": 30,
      "default": 15,
      "description": "The duration of the client-side veto countdown before the delegated trade is committed autonomously."
    }
  },
  "required": ["safety_boundaries", "veto_window_seconds"]
}
\end{lstlisting}
Enforcing this system pattern at the API layer allows collaborative toolkits to resolve the control--delegation tradeoff, giving users the comfort of structural safety constraints (resolving algorithm aversion) while capturing the collective efficiency gains of autonomous model execution.

\paragraph{\rev{Programmatic Two-Pass Guardrail Assertion Layer.}} \rev{Because LLMs are susceptible to prompt injection or semantic alignment decay, unstructured \texttt{strategic\_directives} (e.g., a prompt string instructing the agent to ``be highly cooperative and close trades rapidly'') could theoretically lead the model to generate offers that violate the hard parameters of the \texttt{safety\_boundaries}. To guarantee that these safety bounds are mathematically inviolable, collaborative platforms must implement a programmatic, two-pass guardrail assertion layer.} 

\rev{The systems engine must intercept the model's structured JSON output, parsing and validating the proposed variables programmatically on the backend *before* executing or broadcasting the trade. If the LLM-generated offer fails any of the hard parameters, the transaction is blocked, and a system exception is thrown:}
\begin{lstlisting}
// Programmatic Backend Guardrail Assertion Check
function validateDelegatedProposal(generatedProposal, safetyBoundaries) {
  // 1. Verify Individual Surplus Threshold
  if (generatedProposal.calculated_surplus < safetyBoundaries.minimum_expected_surplus) {
    throw new GuardrailException("LLM proposed trade violates minimum surplus threshold.");
  }
  // 2. Verify Trade Volume Budget
  if (generatedProposal.offer_to_give.quantity > safetyBoundaries.maximum_quantity_per_trade) {
    throw new GuardrailException("LLM proposed trade exceeds maximum chip volume limit.");
  }
  // 3. Verify Excluded Assets Color Blacklist
  if (safetyBoundaries.excluded_assets.includes(generatedProposal.offer_to_give.color)) {
    throw new GuardrailException("LLM proposed trade attempted to offer a blacklisted asset.");
  }
  return true; // Safe to execute/render
}
\end{lstlisting}
\rev{By implementing this programmatic validation step, collaborative toolkits achieve robust, provable system safety, rendering AI delegation suitable for high-stakes real-world organizational workflows.}

\end{document}